\documentclass[sn-mathphys,Numbered]{sn-jnl}

\usepackage{graphicx}%
\usepackage{multirow}%
\usepackage{amsmath,amssymb,amsfonts}%
\usepackage{amsthm}%
\usepackage{mathrsfs}%
\usepackage[title]{appendix}%
\usepackage{xcolor}%
\usepackage{textcomp}%
\usepackage{manyfoot}%
\usepackage{booktabs}%
\usepackage{algorithm}%
\usepackage{algorithmicx}%
\usepackage{algpseudocode}%
\usepackage{listings}%
\usepackage{longtable}
\usepackage{array}
\usepackage{graphicx}
\usepackage{url}
\usepackage{hyperref}
\usepackage{subcaption}

\newcommand*\rotvertical{\rotatebox{90}}

\theoremstyle{thmstyleone}%
\theoremstyle{thmstyletwo}%

\theoremstyle{thmstylethree}%

\begin{document}

\title[]{Testing autonomous vehicles and AI: perspectives and challenges from cybersecurity, transparency, robustness and fairness}



\author*[1,2]{\fnm{David} \sur{Fernández Llorca}}\email{david.fernandez-llorca@ec.europa.eu}

\author[3]{\fnm{Ronan} \sur{Hamon}}

\author[3]{\fnm{Henrik} \sur{Junklewitz}}

\author[4]{\fnm{Kathrin} \sur{Grosse}}

\author[5]{\fnm{Lars} \sur{Kunze}}

\author[6]{\fnm{Patrick} \sur{Seiniger}}

\author[7]{\fnm{Robert} \sur{Swaim}}

\author[8]{\fnm{Nick} \sur{Reed}}

\author[4]{\fnm{Alexandre} \sur{Alahi}}

\author[1]{\fnm{Emilia} \sur{Gómez}}

\author[3]{\fnm{Ignacio} \sur{Sánchez}}

\author*[3]{\fnm{Akos} \sur{Kriston}}\email{akos.kriston@ec.europa.eu}

\affil*[1]{\orgname{European Commission, Joint Research Centre}, \orgaddress{\city{Seville}, \country{Spain}}}

\affil[2]{\orgdiv{Computer Engineering Department}, \orgname{University of Alcalá}, \orgaddress{\country{Spain}}}

\affil[3]{\orgname{European Commission, Joint Research Centre}, \orgaddress{\city{Ispra}, \country{Italy}}}

\affil[4]{\orgname{EPFL - Swiss Federal Institute of Technology}, \orgaddress{\city{Lausanne}, \country{Switzerland}}}

\affil[5]{\orgname{University of Oxford}, \orgaddress{\city{Oxford}, \country{UK}}}

\affil[6]{\orgname{BASt - German Federal Highway Research Institute}, \orgaddress{\country{Germany}}}

\affil[7]{\orgname{Founder NTSB. HowiItBroke.com}, \orgaddress{\city{Washinton DC}, \country{US}}}

\affil[8]{\orgname{Reed Mobility}, \orgaddress{\city{Wokingham}, \country{UK}}}



\abstract{
This study explores the complexities of integrating Artificial Intelligence (AI) into Autonomous Vehicles (AVs), examining the challenges introduced by AI components and the impact on testing procedures,  focusing on some of the essential requirements for trustworthy AI. Topics addressed include the role of AI at various operational layers of AVs, the implications of the EU's AI Act on AVs, and the need for new testing methodologies for Advanced Driver Assistance Systems (ADAS) and Automated Driving Systems (ADS). The study also provides a detailed analysis on the importance of cybersecurity audits, the need for explainability in AI decision-making processes and protocols for assessing the robustness and ethical behaviour of predictive systems in AVs. The paper identifies significant challenges and suggests future directions for research and development of AI in AV technology, highlighting the need for multidisciplinary expertise.
}




\maketitle

\section{Introduction}\label{sec1}
Artificial Intelligence (AI) plays a critical role in the advancement of autonomous driving. It is likely the main facilitator of high levels of automation, as there are certain technical issues that only seem to be resolvable through advanced AI systems, particularly those based on machine learning. However, the introduction of AI systems in the realm of driver assistance systems and automated driving systems creates new uncertainties due to specific characteristics of AI that make it a distinct technology from traditional systems developed in the field of motor vehicles. Some of these characteristics include unpredictability, opacity, self and continuous learning and lack of causality \cite{liabAI}, among other horizontal features such as autonomy, complexity, overfitting and bias. As an example of the specificity that the introduction of AI systems in vehicles entails, the UNECE’s Working Party on Automated/Autonomous and Connected Vehicles (GRVA) has been specifically discussing the impact of AI on vehicle regulations since 2020 \cite{GRVA-AI2022}. 

In order to maximize the benefits of using AI and reduce potential negative impacts, several frameworks have been developed that establish principles and requirements for the development of trustworthy AI. For instance, the OECD AI value-based principles for responsible stewardship of trustworthy AI \cite{OECD2023}, or the EU’s ethical guidelines for trustworthy AI \cite{AIHLEG2019}, \cite{AIHLEG2020}, \cite{EthicsCAV2000} which encompass multiple ethical principles, requirements, and criteria to ensure that AI systems are designed following a human-centered approach. 

These frameworks comprise multiple interrelated components that extend beyond safety concerns, such as cybersecurity, robustness, fairness, transparency, privacy, accountability, societal and environmental well-being, among others. Their general applicability to the autonomous driving domain has been previously analyzed in \cite{Llorca2021}, \cite{Llorca2023}, demonstrating that these are complex frameworks which require addressing multiple problems of varying nature. Some of these problems are still at an early stage of scientific and technological maturity, presenting new research and development challenges in various areas. Progress in this context requires multidisciplinary expertise and tailored analyses.

In this work, based on the methodology described in Section \ref{sec1-method}, we present a detailed study of the state of the art in relation to some of the most prominent elements of trustworthy AI frameworks, and how they affect the current and future landscape of testing procedures for AVs. Specifically, we focus on cybersecurity, transparency, robustness, and fairness. The analysis is conducted by a multidisciplinary group of experts and includes the identification of the most pressing future challenges.

After describing the methodology and presenting the terminology used, Section \ref{sec2},  is dedicated to presenting the various operational layers involved in vehicle automation, as well as the impact of AI on each of them. It also presents the regulatory context of AI and its sectoral application to the field of AVs. Section \ref{sec3} presents the current landscape of vehicle regulation and standards, including ex-ante, post-hoc, and accident investigation processes. Sections \ref{sec4}, \ref{sec5}, and \ref{sec6} detail the requirements for cybersecurity, transparency through explainability, robustness of prediction systems, and fairness. Finally, Section \ref{sec7} presents the general conclusions and future work.

\subsection{Methodology}
\label{sec1-method}
To tackle the main focus areas of our analysis (cybersecurity, transparency, robustness, and fairness), we employed an expert opinion-based methodology at two levels. First, we conducted an interdisciplinary workshop involving 21 expert academics from various disciplines relevant to trustworthy AI and AVs. This interactive, collaborative online workshop allowed each expert to introduce a topic related to one of the focus areas, address all related questions, and provide additional resources as needed. The workshop's outcome was encapsulated in a technical report~\cite{Kriston2023}. From this, a smaller group of experts was selected to go deeper into each focus area. This group aimed to provide a state-of-the-art overview for each area within the context of autonomous driving, specifically focusing on testing methodologies, and subsequently identify the most significant challenges in each area. 

Due to the multidisciplinary nature of trustworthy AI for AVs and the diverse group of experts, it was initially agreed to establish a common terminology, identify the main operational layers of AVs, and analyse the impact of AI on each of them. The regulatory context of AI and vehicles was also collectively examined.

\subsection{Terminology}
With respect to the terminology, in this work, we follow the proposal presented in \cite{Llorca2023}. We use \emph{advanced driver assistance systems} (ADAS) or \emph{assisted vehicle/driving} for SAE levels 1 and 2 (driver), \emph{automated vehicle/driving} for SAE level 3 (a backup driver/user is in charge), and \emph{autonomous vehicle/driving} for SAE levels 4 and 5 (passenger/unoccupied). In some cases, we also use \emph{automated driving systems} (ADS) to generically refer to SAE levels 3 to 5 (automated and autonomous driving). Finally, when we use the acronym AV, we refer to automated and/or autonomous vehicles indistinctly.

\section{AI for AVs}\label{sec2}

\subsection{Operational layers of AVs and the role of AI}\label{subsec3-1}

Autonomous vehicles (AVs) are complex systems that interact with highly complex environments with an almost infinite variety of possibilities. High levels of automation are facilitated by at least seven operational layers (see Figure \ref{fig:layers}):

\begin{figure}[t]
\centering
\includegraphics[width=0.9\textwidth]{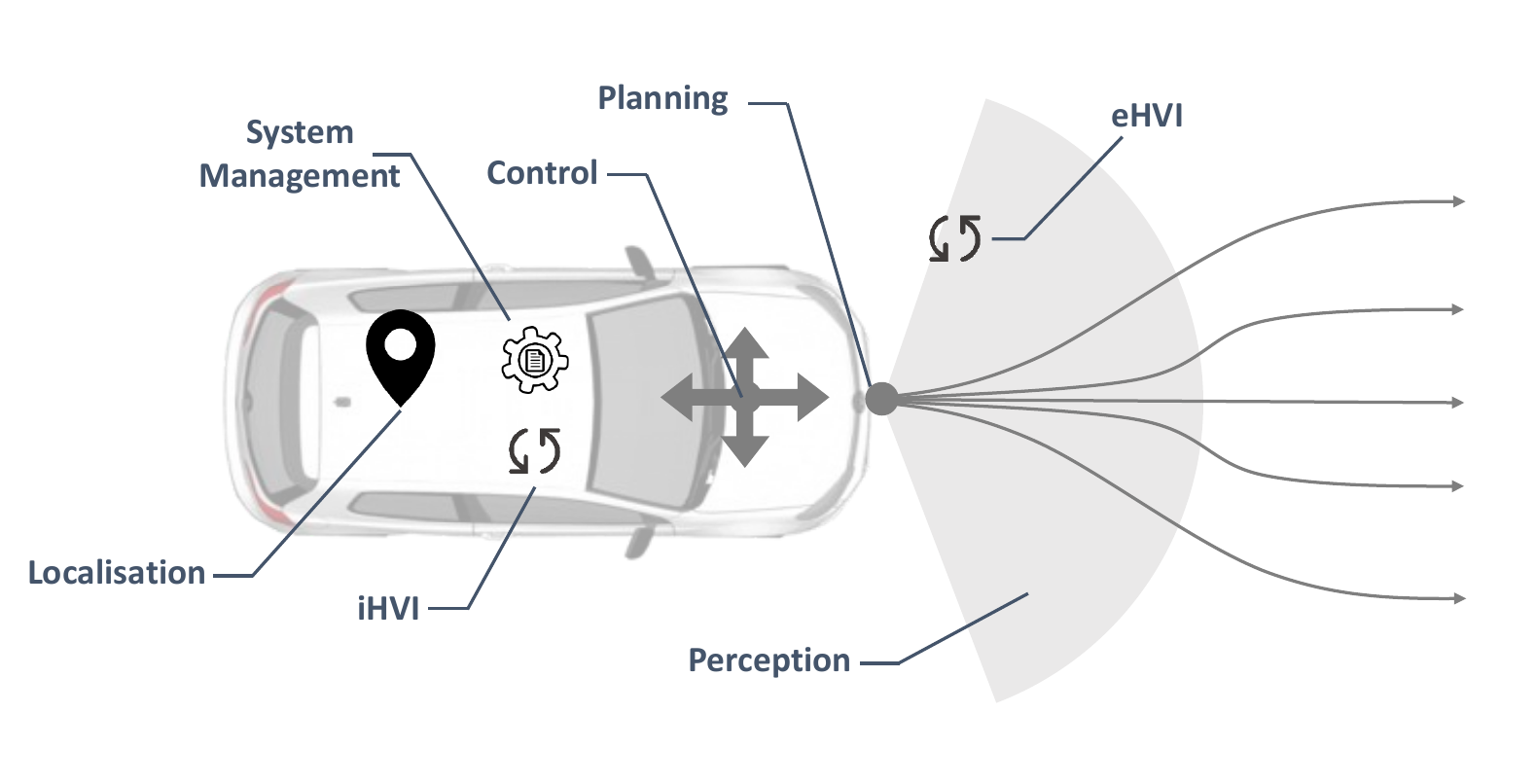}
\caption{Key operational layers: \textit{localisation}, \textit{perception}, \textit{planning}, \textit{control}, \textit{human-vehicle interaction} and \textit{system management}. Adapted and expanded from~\cite{jo2014development}.}
\label{fig:layers}
\end{figure}

\begin{figure}[t]
\centering
\includegraphics[width=1\textwidth]{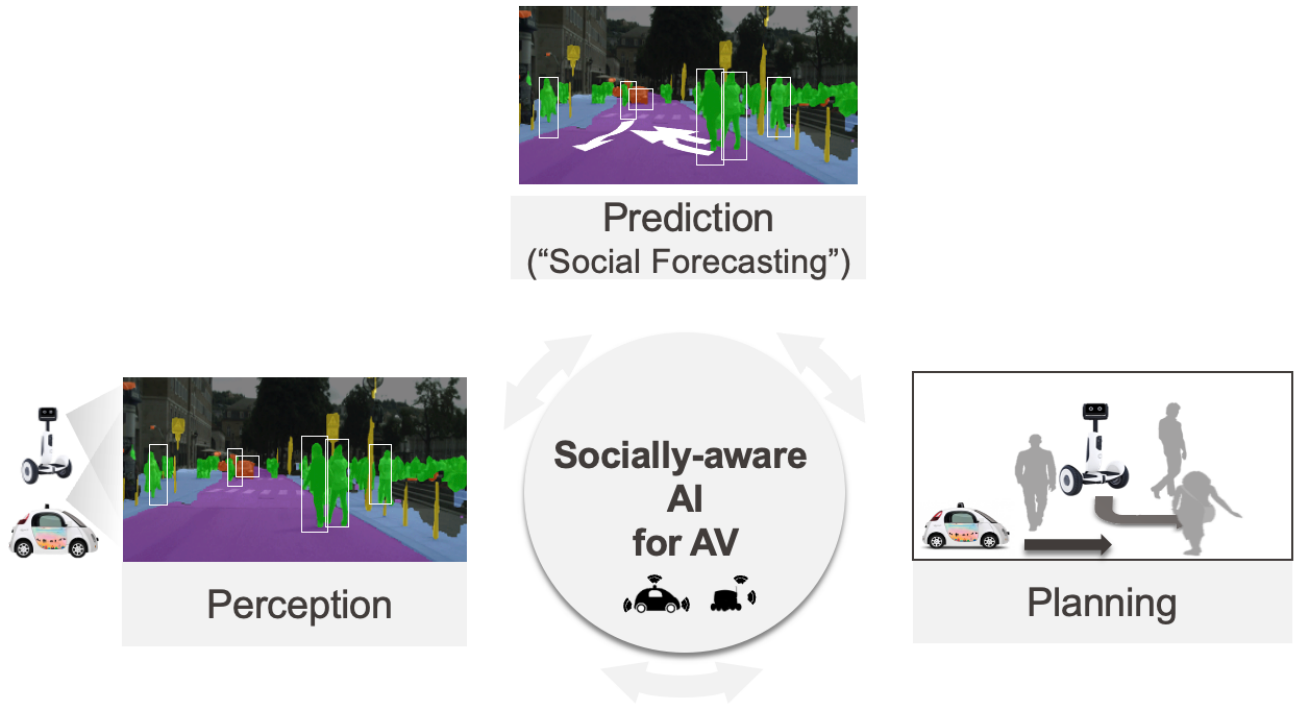}
\caption{Socially-aware AI for AVs. \textbf{\textit{Perception}, \textit{Prediction}, and \textit{Planning}}, also referred to as the \textbf{3 Ps}. \textit{Perception} and \textit{planning} augmented with \textit{prediction} to safely yet effectively deploy AV that will interact with other \textit{social agents}.}
\label{fig:3Ps}
\end{figure}

\begin{itemize}
    \item[] \textit{\textbf{Localisation}}: determines vehicle pose (position and orientation) relative to a digital map. This includes enhanced digital or high-definition maps that contain static elements (e.g., traffic lights, traffic signs, pedestrian crossings, lanes, street layout, intersections, roundabouts, parking areas, green areas, etc.) and dynamic real-time information (e.g., weather and traffic conditions). Different map-matching techniques are used to accurately establish the vehicle's position and heading within the map. While traditionally localisation was not based on learning, data-driven and machine learning approaches are becoming prevalent~\cite{Ma2020}.
    \newline


\item[]\textit{\textbf{Perception (and prediction)}}: a.k.a. scene understanding, this process detects and interprets the dynamic content of the scene, including vehicles and Vulnerable Road Users (VRUs)~ \cite{Llorca2021}. It also includes segmenting the static elements of the scene from the perspective of the vehicle. Perception also includes \textit{prediction} of future behaviours and trajectories of other agents, such as vehicles and VRUs, which are crucial for safe and comfortable planning. \textit{Prediction}, also referred to as \textit{social forecasting}, extends beyond simple predictions to forecast long-term dependencies and complex collective interactions. It models social etiquette, the unwritten rules of social interactions, predicting social agents' behaviours based on social models and including awareness of agents' own actions on others' movements. Social agent interaction requires joint reasoning, and the most promising approaches are based on complex machine learning methods (e.g., socially-aware AI for AVs, as depicted in Figure \ref{fig:3Ps}).
\newline


\item[]\textit{\textbf{Planning}}: a.k.a. (local) motion or trajectory planning or decision making. Based on vehicle localization, knowledge of traffic rules, state of signals, and predicted trajectories of dynamic agents, local planning is performed, providing a feasible and smooth trajectory and speed references for the control system. We can distinguish between different methods that have traditionally been applied to deal with motion planning, including graph-search, variational or optimization-based, incremental or sample-based, and interpolation-based~\cite{Paden2016}. End-to-end deep learning approaches are becoming predominant~\cite{Aradi2022}.
\newline
\item[]\textit{\textbf{Control}}: 
it involves both longitudinal (acceleration and breaking) and lateral control (steering) of an AV~\cite{khodayari2010historical}. Feedback controllers are used to select the most appropriate actuator inputs to perform the planned local trajectory. Many different types of closed loop controllers have been proposed for executing the reference motions provided by the path planning system \cite{Paden2016}, including path stabilization, trajectory tracking, and more recently, predictive control approaches.
\newline
\item[]\textit{\textbf{Human-Vehicle Interaction (HVI)}}: involves the design of human-vehicle interfaces for effective interaction and communication with in-vehicle users and external road users~\cite{Llorca2021}. Potential modalities to communicate intention of the AV to road users include explicit, such as audio and video signals, and implicit forms, such as vehicle's motion pattern (speed, distance and time gap) \cite{Rasouli2020}. Regarding driver and passengers, common interfaces are audio, tactile, visual, vibro-tactile, and more recently, natural language processing \cite{Roh2020}. In addition, this layer also includes in-vehicle perception systems to detect the users’ status.
\newline
\item[]\textit{\textbf{System Management}}: it supervises the overall ADS, including functions such as fault management, data management and logging, event data recording, etc. 
\end{itemize}

Although with varying levels of intensity, we can state that AI plays a predominant role in the operational layers of localisation, perception, planning, and human-vehicle interaction, somewhat less in the control layer, and circumstantially in the system management layer.


\subsection{The AI Act and its sectoral application}\label{subsec3-2}
In April 2021, the European Commission proposed the AI Act \cite{AIAct21}, which aims to establish trustworthy AI practices in the European Union. The AI Act adopts a risk-based approach that mandates specific obligations for providers of AI systems depending on their risk level. High-risk AI system providers (including those with potential risks to fundamental rights, health, and safety of humans) are obligated to meet defined requirements in Title III, Chapter 2 of the legal text (Articles 8 to 15). These requirements refer to risk management, data governance, technical documentation, record-keeping, transparency and provision of information to users, human oversight, accuracy, robustness, and cybersecurity.

Article 6 establishes the rules for the classification of high-risk AI systems. The proposal defines a specific set of AI systems that shall be considered high-risk (Annex III). This can be considered the core of the regulation.  However, the AI Act also affects sector specific regulations. More specifically, AI systems intended to be used as a \emph{safety component} of a product, or as a product itself, covered by Union harmonisation legislation (listed in Annex II, including machinery, toys, medical devices, motor vehicles regulations, etc.), as well as those products requiring a third-party conformity assessment before being put on the market or putting into service. The Annex II includes Regulation (EU) 2018/8584~\cite{2018-858} on the approval and market surveillance of motor vehicles, and amending Regulation (EU) 2019/2144~\cite{2019-2144} on type-approval requirement for motor vehicles as regards their general safety and the protection of vehicle occupants and vulnerable road users. It is important to note that specific requirements for ADAS and ADS are further developed through delegated and implementing acts. 

\subsection{Safety components with AI in AVs}\label{subsec3-3}
In order to be subject to possible future requirements under a Commission delegated act, the AI system supplied or integrated in the AVs should be considered a safety component within the meaning of the AI Act. 

The concept of “safety component” is defined in the AI Act~\cite{AIAct21} as “\textit{a component of a product or of a system which fulfils a safety function for that product or system or the failure or malfunctioning of which endangers the health and safety of persons or property}”. This general definition does not necessarily overlaps with sector-specific definitions of safety components of products covered by Union harmonisation legislation. Particularly relevant is the new Machinery Regulation (EU) 2023/1230 \cite{2023-1230}, in which the concept of safety component explicitly specifies that the component ``\textit{is not necessary in order for that product to function or for which normal components may be substituted in order for that product to function}". Then, an indicative list of safety components is provided in Annex II, including presence detectors of persons, monitoring devices, emergency stop devices, two hand control devices, among others. Another example is Directive 2014/33/EU \cite{2014-33-EU} relating to lifts and safety components for lifts, which also includes a list of safety components well aligned with the idea of “\textit{not necessary for the product to function}”, such as locking landing doors devices, fall prevention devices, overspeed limitation devices, etc. 

As stated in \cite{Llorca2021, Lolfing2023}, given the inherently high-risk nature of AVs (capable of causing serious physical harm or death, as well as property damage), it is arguably quite intuitive to qualify AI systems of all the operational layers of AVs as safety components. All functions addressed by the different operational layers fulfil a safety function that may endanger the health or safety of persons or property in case of malfunction. The point is that, at the same time, these functions are essential for ``the product to function". We can identify some components that are not necessary for the overall functioning of the system, such as comfort functions, infotainment applications, or vehicle warnings. However, these functions are not safety relevant either. In other words, virtually all components of AVs that are safety relevant are also relevant to the overall functioning of the system. Therefore, the conventional definition of safety component given in contexts such as machinery or lifts may not be applicable to AVs. In any case, considering that an average passenger car contains more than 30000 parts, among which there will be a wide variety of AI systems for different functions (with a tendency to increase), at some point it should be clearly defined which AI systems used in which specific context can be considered as safety components within the meaning of the AI Act.

\section{Vehicle regulations and standards}\label{sec3}

\subsection{Ex-ante requirements and testing}\label{subsec2-1}
The product development process of any given system, including ADAS and ADS, is typically based on the V-model of product development (e.g., ISO 26262). Over the course of the development of a system, the degree of abstraction of the system properties first decreases (in the left branch for development) and then increases again (in the right branch for verification). At the same time, the objective shifts from questions of system design (left branch) to validation of the intended system characteristics (right branch). Requirements from vehicle technical regulations or standards and guidelines can also be placed in this scheme, giving vehicle safety bodies a chance to directly influence the development process \cite{Martin2017}.


External requirements can come from various sources, most prominently the type-approval requirements, and independent car assessment programs such as the European New Car Assessment Program (Euro NCAP) \cite{EuroNCAP2022}. Euro NCAP procures and tests market-available vehicles (compliant with type-approval requirements), comparing performance via standardized procedures. As these procedures are communicated in advance, manufacturers can integrate anticipated test results into their product development cycles.

All vehicles registered in Europe adhere to the EU Type Approval Directive 2007/46/EC \cite{2007-46-ec} or the EU Type Approval Regulation (EU) 2018/858 \cite{2018-858}, including amendments like the General Safety Regulations (GSR) Regulation (EC) No. 2019/2144 \cite{2019-2144}. However, these revised directives and regulations do not specify technical requirements but refer to delegated or implementing acts of the European Commission (e.g., Implementing Regulation (EU) 2022/1426 regarding ADS or fully automated vehicles \cite{2022-1426}) or designated UN regulations (e.g., UN Regulation No 157 on Automated Lane Keeping Systems \cite{ECE/TRANS/WP.29/2020/81}). These are preferred as they allow global harmonization of the requirements. The amended documents list mandatory or optional requirements for vehicle type approval. Thus, UN regulations increasingly impact ADAS and ADS. In fact, ADS can serve as an example of optional regulations (so-called "if fitted" regulations). The framework regulation also allows exemptions for new technologies or concepts (see Art. 39 in \cite{2018-858}), enabling deviation from defined requirements to foster innovation, given that road safety and environmental protection are not compromised.


It is important to highlight the difference between conventional and new regulatory approaches. Conventional regulations typically set implicit requirements, encompassing a test procedure with few conditions and defining pass criteria solely for those conditions. Conversely, newer regulations, particularly for ADAS and ADS (see \cite{2022-1426}), define requirements across a range of operating conditions without limiting potential test cases. They also involve "market surveillance" processes that allow regulators, namely the European Commission and Member States, to retest production vehicles against these broad requirements. Therefore, vehicle manufacturers are required to design robust systems capable of functioning under a variety of circumstances. 


Conventional testing for ADAS and ADS employs various methods contingent on the test objective, such as the system functionality to be evaluated and the development status. The methods typically define the general procedure, tools (e.g., artificial target objects, measurement and control devices, suitable environment), procedures for deriving characteristic values from collected data (e.g., calculation of speed reduction in an emergency braking system), and often include an evaluation standard for deriving a test result from the characteristic values (e.g., weighting table depending on test speed in Euro NCAP emergency braking assistance tests) or reference to such (e.g., referencing requirements of UN Regulation no. 152 \cite{ECE/TRANS/WP.29/2020/9}). If assessment standards are specified, they can be interpreted as implicit requirements for the corresponding system.


Advance testing procedures, e.g., for evaluating ADAS with environment perception or ADS, require generating relevant application situation for each system. As public road testing is often unfeasible, equivalent situations must be created in a test environment. In the case of tests for emergency reactions of ADS, for example, this requires objects that represent stationary or moving vehicles. Distinguishing features of test procedures may include the type of these target objects, their motion systems, or the degree of virtualization. 


Test procedures for ADAS and ADS primarly involve position measurement, position control with driving robots (e.g., automated vehicles control their position by themselves) and movable targets. The state-of-the art for position measurement are inertial systems with stabilization by differential GNSS (e.g., GPS, GLONASS), capable of measuring vehicle position with 1 cm accuracy. Derivative and angular accuracy are also reasonably accurate.


For ADAS or ADS that require driver input, the "driving robot systems" can manipulate the controls. These systems can perform open-loop input into the controls and control vehicle movement to follow pre-programmed trajectories. Regarding target systems, it is common to use  dummies emulating cars, bicycles, motorcycles or pedestrians for ADAS or ADS testing. These targets, standardized per ISO 19206, are typically propelled by self-driving, overrunable robot platforms.


However, there are significant limitations in current testing methodologies for ex-ante requirements validation. Test procedures must sufficiently represent or simulate all necessary information for each situation, increasing the effort needed to develop and represent realistic and relevant test situations for ADAS and ADS. It is also necessary to prove that these situations accurately represent reality for a system response. For instance, used objects must not only be sensor-detectable but also representative of vehicles, pedestrians, or other real traffic road users. Without this proof, despite excellent functional and safety evaluations under artificial conditions, little safety progress is achieved in real road traffic.


The same issue applies to the selection of test scenarios \cite{Menzel2018}. As the number of sensors and parameters involved in the operation of ADAS and ADS increases, the effort to define scenarios, execute tests, and prove field relevance escalates, especially for predictive systems based on agent behavior modeling \cite{Izquierdo2022}. On the other hand, it would still be possible to fine-tune the systems to react to standardized targets, but not to, for example, pedestrians with different clothing and appearance. The primary challenge is verifying that ADAS and ADS function only on test tracks under optimal conditions, but also in realistic environments and with parameters outside the test cycles.

\subsection{Post-market monitoring}\label{subsec2-2}


Market surveillance mechanisms, such as those defined by EU Regulation 2018/858~\cite{2018-858}, ensure that motor vehicles and components on the market comply with laws, regulations, and safety and health requirements. The European Commission participates in these activities, verifying vehicles' emissions and safety compliance, and collaborating with national authorities for information exchange and peer review of technical actors~\cite{Bonnel2022}. While the requirements for market surveillance match those of ex-ante type-approval procedures, the testing methodologies may differ considerably.


Another interesting tool recently integrated in the Implementing Regulation (EU) 2022/1426~\cite{2022-1426} is the "In-Service Monitoring and Reporting" (ISMR), which addresses in-service safety of ADS post-market (i.e., operational experience feedback loop). It is primarily based on in-service data collection to assess whether the ADS continues to be safe when operated on the road and identify safety risks~\cite{GRVA-VMAD2022}. 


The AI Act~\cite{AIAct21} also outlines several post-market actions, including market surveillance, human oversight and monitoring (as a requirement), and post-market monitoring. Post-market monitoring involves all activities carried out by providers of AI systems to collect and review experience gained from the use of AI systems to identify any need to immediately apply corrective or preventive actions. This idea is well aligned with the ISMR concept. 

\subsection{Crash investigation}\label{subsec2-3}

Despite diligent design, planning, and safety regulations, accidents occur. Investigations into root causes identify and address contributing factors to prevent recurrence and/or mitigate negative outcomes. The investigation depth varies, ranging from counting crashes in a timeframe to detailed examinations of involved parties and vehicle operation assumptions. Collisions typically involve a mix of human, machine or system, and environmental factors. When AVs are involved in an accident, investigators must establish the required investigation depth. The need of police officers often end once they file a report recording scene, damage, interviews, and traffic law violations. These facts might be \textit{contributing factors} but at this point, they represent basic background information for examining how an automated system may have failed. Collecting facts and assembling a chronology of events leads to a causal chain establishment. However, "probable cause" determination for crash investigations is not the same as for criminal prosecutions, which require "facts beyond a reasonable doubt". 


The investigation depth is a balance between available resources and required information. All investigations involve fact collection, analysis of how facts fit together, and proceeding forward with new understanding. The causal chain typically involves a mix of \emph{man and machine}. Various methods convey information and establish crash causes, including the "Five M's and E" \cite{5mse-2011}, the "Swiss Cheese Model" cite{Reason2006}, and the "Fault Tree Analysis" method \cite{FTA2002}.
Recorded information and facts are often embedded in multiple vehicle devices and other sources. The recorded data may contain a variety of parameters, requiring manufacturer assistance to obtain and specialized software to interpret. From the investigation and fact collection, a chronology of the accident can be established to identify contributing factors and probable cause. 


With increasing numbers of vehicles with ADAS and ADS, investigators may question how much these automated systems contributed to collisions. Factors complicating this include drivers' lack of understanding of how ADAS/ADS work, blame shifting to the vehicle, proprietary nature of ADAS/ADS designs, increased time required for ADAS/ADS investigations and potential investigator shortage, and lack of ADAS/ADS training for collision investigators. 


While ADAS or ADS in ground vehicles is relatively new, aviation automation has existed for 110 years, maturing into digital system forensics. Despite design and standard differences, the basic elements involved in ground vehicles and aviation remain a continuous loop of requirements, outputs, actuators, and feedback. Human involvement extends to each of these, necessitating context investigation about man and machine involvement in each aspect, as well as potential environmental contributions. 
Aviation has a long history of autopilot use with occasional involvement in accidents. The two main autopilot-related accident causes have been the human interface and the lack of human supervision or misuse.  Two examples are the accidents of the Air France A330 flying from Rio de Janeiro to Paris in 2009 \cite{BEA2009FinalReport} (human interface issue), and the American Airlines Flight 965 flying from Miami to Cali in 1995 \cite{1995AmericanColumbia} (human supervision issue). Similarly, numerous accidents involving ADAS and ADS in road vehicles have occurred when drivers have fallen asleep or intentionally tried to let the vehicle do more than the automation level was intended \cite{2020TeslaCalifornia, 2019CollisionSemitrailer}. 
 
Determining proportionality in an crash investigation has no singular easy method. The process continues to be data collection, beginning at the scene and not starting analysis until factual elements are fully collected. The challenge for ADAS and ADS investigation is that there may be substantially more data to pursue and more unforeseen aspects to learn about, some of which may be proprietary and difficult to obtain.

\section{Cybersecurity audits for AI components}\label{sec4}

Cybersecurity standards have been established for motor vehicles~\cite{CysecCars}, with ADS introducing an additional layer of complexity due to AI components. Existing standards may not necessarily apply to AI~\cite{salay2017analysis}, which has been shown to introduce security vulnerabilities, impacting any system relying on or incorporating AI components. We consider a security vulnerability as the potential for malicious activity that alters the behaviour~\cite{gu2017badnets,xiang2021backdoor,wen2021generative,morgulis2019fooling} or leaks data~\cite{liu2022ml,shokri2017membership,zhu2021hermes, tramer2016stealing} from an AI component, designed for a specific task. To prevent attacks on critical infrastructure, like AVs, we propose an AI security audit structure for standardized assessment of system security against specific threats or resistance to information leakage.


Before discussing an audit structure, we need a risk assessment of likely threat occurrence. We propose four levels of AI security research concreteness. The first level describes theoretical research independent of the application or ADAS/ADS, and the fourth level describes a vulnerability exploited in practice. From this perspective, we review three relevant security threats associated with AI components in ADS: \textit{evasion}, \textit{poisoning}, \textit{privacy} and \textit{IP}. Other cybersecurity risks independent of the AI components will not be discussed here. For an overview, we refer to~\cite{mclachlan2022tempting}.

\subsection{Security threats}
\begin{figure}[t]
\centering
\includegraphics[width=1\textwidth]{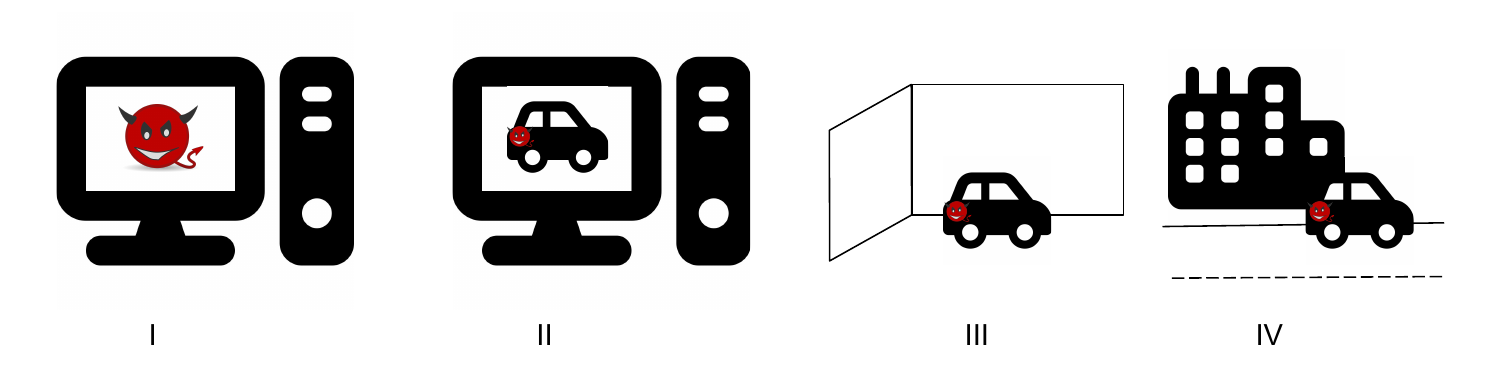}
\caption{Four levels of concreteness in AI component AV security. \textbf{I}, a vulnerability is known to exist for an AI. \textbf{II}, a vulnerability is shown for an AI that is potentially used in cars. \textbf{III}, the vulnerability has been shown on a testbed, or on a car in a safe environment (e.g., driver is aware of the test). \textbf{IV}, the exploit has been used on a car on the road.}
\label{fig:concretenessKathrin}
\end{figure}

This section primarily focuses on classification, vision and LiDAR tasks, due to readily available works in the area and the direct exposure of perception modules to the outside world, potentially increasing vulnerability. In discussing practicality of studying threats, we propose four levels of concreteness related to ADAS and ADS threats (see Figure~\ref{fig:concretenessKathrin}):

\begin{itemize}
    \item[] \textbf{I} A vulnerability demonstrated on an AI algorithm used in an AV on a non-vehicle related task (e.g. face detection).
    \item[] \textbf{II} A vulnerability demonstrated on an AI algorithm used in an AV on a relevant task (e.g. image segmentation or traffic sign classification).
    \item[] \textbf{III} A vulnerability demonstrated on an AI algorithm used in an AV on a relevant task, with experiments showing that the vulnerability is presence in a vehicle or similar testbed.
    \item[] \textbf{IV} A reported vulnerability affecting an on-road AV.
\end{itemize}


While concreteness increases towards real-world threats, a vulnerability can directly occur at level III or IV without being observed at lower levels first. Studying a vulnerability at level I is not a prerequisite for its occurrence at concreteness II, III, or IV. 

It is also relevant to distinguish between \emph{adversarial} and \emph{benign accuracy} when considering AI component vulnerability in AVs. Both are based on accuracy, measured on a specific dataset called the test set, while the AI is optimized on a separate dataset, the training data. Adversarial accuracy is the accuracy when an attacker targets the system to decrease performance using a vulnerability by altering training or test data, or both. Benign accuracy is the accuracy when no attacker is present, though performance decreases may still occur, for example, when test and training data differ. This may happen when a car is released for production and operated in real-world traffic as opposed to recorded scenes, where unknown weather conditions may occur, or unanticipated car behaviour may be observed.

\subsubsection{Evading AVs}

In evasion attacks, an attacker alters inputs at test time. These modified inputs may appear as one class to a human observer but get misclassified by the model despite otherwise good performance. For example, a traffic sign recognition algorithm might misclassify a manipulated “stop sign” image as a “yield sign” assignment~\cite{evtimov2017robust}. Such manipulations can be achieved through classifier-specific graffiti~\cite{evtimov2017robust} or subtle changes to the image~\cite{kurakin2018adversarial} or sign shape~\cite{tu2020physically,evtimov2017robust}. Evasion is a vulnerability that can occur at deployment or during testing stages and exists across all types of classifiers~\cite{papernot2016transferability} and input data or tasks~\cite{evtimov2017robust,tu2020physically,vemprala2021adversarial}, often referred to as adversarial examples. AV-specific evasion attacks can target visual sensors~\cite{evtimov2017robust} or other sensors like LiDAR~\cite{tu2020physically}. They can also directly target the vehicle's planner or affect end-to-end driving by adding shadows or lines on the road~\cite{boloor2019simple}. Most attacks proposed on AVs focus on one component and do not trigger complex behaviours. We attempt to give a concise overview of suggested attacks in Table~\ref{tab::evasionAttacksKathrin}. Real-world incidents with vision systems~\cite{mcgregor2021preventing,grosse2022so} are generally assumed to be benign incidents, not attacks by a malicious entity. 

\begin{table}[htp]
\caption{Selection of evasion attacks on AVs with different levels of concreteness as depicted in Figure~\ref{fig:concretenessKathrin}.}\label{tab::evasionAttacksKathrin}
\centering
\begin{tabular}{p{0.05\linewidth}p{0.14\linewidth}p{0.18\linewidth}p{0.15\linewidth}p{0.3\linewidth}}
\toprule
\textbf{Ref.} & \textbf{Target component} & \textbf{Target sensor} & \textbf{Concreteness} & \textbf{Description} \\
\midrule
 \cite{evtimov2017robust} & Downstream & Camera & II & Grafiti leading to wrong classification \\
\cite{tu2020physically} & Downstream & LiDAR  & II & Adds objection to hide object  \\
\cite{boloor2019simple} & End-to-end & Camera & II & Adding lines on street lead to wrong steering output \\
\cite{vemprala2021adversarial} & Planner & Motion sensors & II &  Alter movement feedback causes wrong trajectory \\
\cite{morgulis2019fooling} & Downstream & Camera & III & Perturbing traffic signs \\
 \cite{grosse2022so} & Unknown & Camera  & III-IV & 2 reported benign incidents \\
\cite{mcgregor2021preventing} & Unknown & Unknown &  IV & Tesla crashes with truck, cause probably benign \\
\bottomrule
\end{tabular}
\end{table}


Many AI engineers and practitioners consider evasion or adversarial examples more of a performance issue than a security problem~\cite{grosse2022so}. Traditional security vulnerabilities (i.e., non-AI related) seem to be more prevalent in practice~\cite{grosse2022so}. However, defence and forensics against evasion attacks remain challenging, making it essential to avoid situations where such attacks do not lead to penalties. 



Defending against evasion is complex as adversarial examples are inherent to the AI functionality~\cite{ilyas2019adversarial}, not a bug. While numerous defences have been introduced, many provide a false sense of security~\cite{carlini2019evaluating} as they prevent specific attacks rather than general adversarial examples. Current standards include adversarial training and formal verification~\cite{pulina2010abstraction} for small critical applications to obtain a safe network. Both approaches must be carried out during development or before deployment by the model provider. However, integrating vulnerability fixes upfront is always preferred over ad-hoc solutions post-design.

\subsubsection{Poisoning of AVs}

In contrast to the latter backdoor attacks, there are few poisoning attacks on systems used in AVs. The reason for this inherent complexity of poisoning attacks is that, in contrast to for example evasion attacks (that derive a change against a static, deployed AI), poisoning attacks aim to make the AI deviate in a malicious way during training, where the AI is still dynamically adapting to the data~\cite{munoz2017towards}. Some works exist on deep learning~\cite{munoz2017towards, jiang2020poisoning}, where~\cite{jiang2020poisoning} tackle traffic sign recognition as used in AVs. However, also the path finding of the vehicle can be poisoned to only find a sub-optimal route~\cite{ma2019policy}.



In poisoning attacks, an attacker alters training data, which affects test time performance. They can degrade data quality~\cite{jiang2020poisoning} or install a backdoor, causing the model to misclassify inputs when a specific pattern or trigger is present~\cite{gu2017badnets}. These attacks are particularly severe as affected models perform well as long as the trigger is not present~\cite{gu2017badnets,wang2021stop}. 

Despite the complexity when compared with evasion, there are several examples. Some works exist on perception~\cite{munoz2017towards, jiang2020poisoning} and planning ~\cite{ma2019policy} operational layers. However, backdoor attacks are easier to implement by controlling part of the training data~\cite{munoz2017towards}. They have been tested on vision models~\cite{gu2017badnets}, image segmentation~\cite{han2022physical}, LiDAR sensors~\cite{xiang2021backdoor,wen2021generative}, and even on traffic congestion control systems~\cite{wang2021stop}. Most of these attacks focus on ADS but do not target real cars, as we show in Table~\ref{tab::poisoningAttacksKathrin}. There are no known incidents of poisoning attacks from media or prior research. 

\begin{table}[htp]
\caption{Selection of poisoning/backdooring attacks on AI components of AVs with their concreteness. }\label{tab::poisoningAttacksKathrin}
\centering
\begin{tabular}{p{0.07\linewidth}p{0.16\linewidth}p{0.09\linewidth}p{0.11\linewidth}p{0.14\linewidth}p{0.23\linewidth}}
\toprule
\textbf{Ref.} & \textbf{Target component} & \textbf{Target sensor} & \textbf{Backdoor?} & \textbf{Concreteness} & \textbf{Description} \\
\midrule
\cite{jiang2020poisoning} & Downstream & Camera & & II & Reduce accuracy of traffic sign recognition \\ 
\cite{gu2017badnets} & Downstream & Camera & \checkmark & II & Misclassify traffic signs with trigger \\
\cite{xiang2021backdoor,wen2021generative} & Downstream & LiDAR & \checkmark & II & Add trigger to missclassify object \\
\cite{ma2019policy} & Planner & Position & & II & Poisoning for suboptimal route choice \\
\cite{wang2021stop} & Traffic control & Position & \checkmark & II &  Add trigger to cause traffic jam \\
\cite{han2022physical}$^*$ & Downstream & Camera & \checkmark & III & Leads to wrong lane detection output \\
& \multicolumn{3}{l}{\footnotesize{$^*$tested on a small testbed vehicle.}}\\
\bottomrule
\end{tabular}
\end{table}

When discussing real-world attack impacts, we must differentiate between poisoning and backdoor attacks~\cite{grosse2022so}. Poisoning attacks significantly affect model performance, making it unlikely that a poisoned model ends up in production. However, backdoors represent a more severe vulnerability, as the high benign accuracy and unknown triggers make the attack easy to overlook.

Defending against poisoning is somewhat understood, with known trade-offs linking attack strength with detectability by the defender~\cite{frederickson2018attack}. However, defending against backdoors remains an open problem for different modalities~\cite{shokri2020bypassing, xiang2022detecting}. In relation to audits, two different approaches exist: preventing a backdoor from being embedded in the network during training~\cite{cina202wild}, and identifying whether a network is backdoored post-training~\cite{cina2023wild}. Both attacks stem from the data collection stage, allowing for countermeasures like data cleaning before training or AI protection during training~\cite{cina2023wild}. Fixing vulnerabilities should be part of a system design approach taken upfront, not done in an ad-hoc manner.

\subsubsection{Attacks on privacy and IP in AVs}

Privacy concerns in AVs generally arise in the context of data leakage from an AI model~\cite{shokri2017membership}. Depending on the data used for training (e.g., real world crashes investigations), data leakage might have privacy implications. It is currently unclear whether video sequences can be retrieved from a model in sufficient detail. A more significant concern is when data from different users is combined to train a new AI, e.g, to improve AV usability based on individual cars. This has been shown to be vulnerable to attribute inference~\cite{melis2019exploiting} or to the complete training inputs~\cite{geiping2020inverting}. Independent of the data used, the model can become a target of an attack where it is copied without the owner's consent by observing input and outputs~\cite{tramer2016stealing} or other data flow within hardware components~\cite{zhu2021hermes}. Privacy and IP threats almost uniquely occur during deployment but are caused by design choices early in the development of the AV and its components. While there is a large body of work about privacy in AVs in general~\cite{liu2021machine}, few works investigate either privacy attacks or data protection, or IP attacks/protection related to AIs deployed within cars. An exception is a study that shows that some AI components can be stolen when observing data flow in the Peripheral Component Interconnect (PCI) bus present in AVs~\cite{zhu2021hermes}. 


Given the surge in privacy-related legislation, privacy attacks are a threat. Although there are reports about privacy being leaked by companies\footnote{\href{https://www.smh.com.au/property/news/wake-up-call-for-real-estate-agencies-harcourts-hit-by-data-breach-20221103-p5bvaq.html}{https://www.smh.com.au/property/news/wake-up-call-for-real-estate}}$^{,}$\footnote{\href{https://www.bleepingcomputer.com/news/security/vodafone-italy-discloses-data-breach-after-reseller-hacked/}{https://www.bleepingcomputer.com/news/security/vodafone-italy-discloses-data-breach}} or AI companies\footnote{\url{https://www.pcmag.com/news/report-ai-company-leaks-over-25m-medical-records}}, there are no reports yet about data protection failures or private data directly leaked from AVs. Analogously, there are no reports about stolen models from AVs. 


In the case of these attacks, we need to differentiate between the loss of IP and insensitive training data, and the car user's privacy. One is considered the company's responsibility and need not be audited in a security audit by a third party unless desired by the audited party. Whenever the users' privacy might be at stake, the model needs to be audited thoroughly. 


Neither privacy nor IP protection is well understood or investigated~\cite{liu2022ml,geiping2020inverting} for AI modules in AVs. Techniques like differential privacy or data noising can alleviate some threats~\cite{liu2022ml}. A recent line of work finds that Bayesian methods provide differential privacy by default~\cite{mir2012differentially,wang2015privacy}, and should be used in sensitive settings. In general, the loss of sensitive information can be restricted by providing less information to the attacker, for example, by restricting the number of queries that can be made per time unit or the kind of information released (e.g., less numerical precision).

\subsection{Testing and auditing}

Given the diverse vulnerabilities discussed, the need and utility for audits vary. In the case of evasion, an audit is useful as it evaluates the model's benign accuracy. Adding these tests enables easy later extension against evasion attacks, and auditing benign accuracy indirectly tests for poisoning. Any production model should be audited against existing backdoors. As for privacy, the AV must be audited to prevent leakage of potential users' sensitive information. Lastly, audits related to IP leakage can be made optional, depending on the model owner's wishes, and need not be conducted by a third party.


We suggest high-level audit criteria for all vulnerabilities, which include:

\begin{itemize}
    \item Validating the design of the model.
    \item Testing the robustness of the model within the system.
    \item Testing the soundness of potential defence measures within the system.
\end{itemize}



The first point refers to the correct usage of data, which steps (or defences) are implemented and why, and which methods are chosen. This is particularly relevant concerning privacy or when methods with formal guarantees are relied on. The audit then needs to assess if the formal criteria match the application and make sense within the larger system under consideration. The second and third points are empirical and assess the model's performance via tests. The third point is added for completeness in cases like evasion and might become relevant later if more attacks occur in practice. For other cases, like backdoors, it is already relevant. In the case of privacy, it might also be necessary to empirically validate a defence that does not provide enough formal guarantees.

\subsubsection{Prerequisite for a cybersecurity audit}

In a nutshell, a cybersecurity audit should assess the model's robustness using an appropriate attack (evasion) or assessment (backdoors, privacy). This raises the question of good attacks and assessments, which generally differ in their threat model and complexity, encompassing access to the model, expected changes by the attacker (norms), and the complexity of the evaluation.

\begin{itemize}
    \item[] \textit{\textbf{Model access}}:  whether the attack uses insider knowledge about this model (white-box) or not (black-box). White-box attacks are considered a worst-case scenario and are stronger than black-box attacks. However, they can also give a false sense of security when not properly tailored for the model. In AVs, we can likely assume that the AI components will be proprietary and thus black-box for an attacker (unless there is an IP breach). For audits, white-box access must be given to study a worst-case attacker, which requires the audit to be performed by a trustworthy, independent third party that is provided confidential access to the model.  
    
    \item[] \textit{\textbf{Norms}}: how an attacker may alter a sample reaching the AI component. To measure the change added to a sample, distance or similarity measures are often used. The most basic functions are called $L_p$ norms, where $p$ denotes the type of alteration captured by this specific norm. For instance, the $L_0$-norm results in sparse, graffiti like alterations, whereas the $L_\infty$-norm creates minor changes in an image. While much of the research focuses on these norms for vision, other norms that mimic human perception~\cite{luo2018towards} can also be utilized. For LiDAR sensors, different norms that focus on adding a consistent object are required~\cite{tu2020physically}. The norm needs to be chosen consistently with the task and robustness requirements. However, there is no scientific consensus on a norm-based threat model for autonomous cars that would help determine which norms or similarities can be used for an evaluation.
    
    
    \item[] \textit{\textbf{Audit complexity}}: the average expected run-time of the evaluation. Complete search of vulnerability is possible~\cite{pulina2010abstraction}, but not practical. When choosing a method for auditing, a trade-off between run-time and expected success rate to identify a vulnerability or data leak needs careful consideration. Any corresponding choices should be well documented.
\end{itemize}

Developing new evasion attacks is an active area of research~\cite{croce2020reliable}. Complex attacks are often preferred as they are more likely to find a malicious sample. An important distinction when auditing evasion is whether targeted (e.g., image should be classified as `yield') or untargeted (e.g., output should not be `stop sign') attacks are audited. Targeted attacks, while providing greater insight into model analysis and posing higher potential harm in autonomous driving, are more challenging to execute due to their requirement for proprietary access. They are also less transferable to unseen models. Therefore, both targeted and untargeted attacks should be incorporated in audits and robustness tests.



On the other hand, developing detection methods that provide security on whether a network is backdoored is an open research question~\cite{cina2023wild}. We suggest testing several different state-of-the-art approaches in terms of heuristics and assumptions. Similarly, there is no conclusive result on how to detect or prevent privacy leakage from a model~\cite{liu2022ml}. We suggest testing attacks introduced in research to get an estimate of existing vulnerability to inference.

\subsubsection{Implementing cybersecurity audits}

Implementing audits for a set of attacks or defences necessitates the use of specific libraries and tools, many of which focus on image classification. Table~\ref{tab:librariesKathrin} provides an overview of these, including supported frameworks and their number of implemented attacks and defences. Notably, ART, SecML, and Privacy Meter support more diverse areas. SecML and Microsoft's counterfit\footnote{https://github.com/Azure/counterfit/tree/main/docs/source} allow command-line robustness testing, while ART and foolbox offer plug-in capabilities.

\begin{table*}[h]
\caption{Overview of different libraries implementing different attacks on ML. We skip repositories with less than 5 evasion/poisoning or less than 2 privacy/IP attacks. The activity reports the last month/year where an activity took place as of December 2023.}
\centering
\resizebox{\linewidth}{!}{
\begin{tabular}{lrlrlrlrllllrrrrrrr}
\toprule
& \multicolumn{2}{c}{Evasion} & \multicolumn{2}{c}{Poisoning} & \multicolumn{2}{c}{Privacy/IP} & \multicolumn{6}{c}{Supported frameworks}\\ 
\cmidrule(l){2-3}\cmidrule(l){4-5}\cmidrule(l){6-7}\cmidrule(l){8-13}
Name & \rotvertical{\# Attacks} & \rotvertical{\# Defenses} & \rotvertical{\# Attacks} & \rotvertical{\# Defenses} &\rotvertical{\# Attacks} & \rotvertical{\# Defenses} & \rotvertical{Pytorch} & \rotvertical{Tensorflow} & \rotvertical{JAX} & \rotvertical{MXNet} & \rotvertical{Caffe} & \rotvertical{SciKit} &  \rotvertical{Non-vision} \rotvertical{Support} &\rotvertical{Github} \rotvertical{stars} & \rotvertical{Last}\rotvertical{activity} & \rotvertical{Released?} & License \\
\midrule
ART~\cite{nicolae2018adversarial} & 41 & 29 & 11 & 7 & 16 & - & \checkmark & \checkmark & &\checkmark & & \checkmark  & \checkmark & 4.2k & 12/23 & \checkmark & MIT \\ 
BackdoorBox~\cite{li2023backdoorbox}& - & - & 12 & 7 & - & - & \checkmark & &&&&& ? & 325 & 10/23 & & GPL 2.0\\
Foolbox\cite{rauber2017foolbox} & 17 & - & - & -& -& -& \checkmark & \checkmark & \checkmark & & & &  & 2.6k & 11/23 & \checkmark & MIT \\ 
SecML~\cite{pintor2022secml} & 10 & 1 & 3 & - & - & - & \checkmark & \checkmark & & & & \checkmark & \checkmark & 126 & 5/23 & \checkmark & Apache2.0 \\ 
AdvBox~\cite{goodman2020advbox} & 10 & 6 & 2 & - & -& - & \checkmark & \checkmark & & & \checkmark & & & 1.3k & 8/22 & & Apache2.0\\ 
BackdoorBench~\cite{wu2022backdoorbench}& - & - & 12 & 15 & - & - & \checkmark & &&&&&  & 249 & 12/23 & & CC BY-NC-4.0\\
Cleverhans~\cite{papernot2018cleverhans} & 6 & - & - & - & - & - &  \checkmark & \checkmark & \checkmark & &  & & & 6k & 1/23 & \checkmark & MIT \\
ML-Doctor~\cite{liu2022ml} & - & - & - & - & 6 &- & \checkmark & & & & & & & 75 & 12/23 & & Apache2.0  \\
Privacy Meter\footnote{\url{https://github.com/privacytrustlab/ml_privacy_meter}} & - & - & - & - & 7 &- & \checkmark & \checkmark & & & & & \checkmark & 501 & 07/23 & & MIT \\
ML Hospital\footnote{\url{https://github.com/TrustAIResearch/MLHospital}} & - & - & - & - & 6 & 10 & \checkmark & & & & & & & 39 & 04/23 & & MIT\\
\bottomrule
\end{tabular}
}
\label{tab:librariesKathrin}
\end{table*}



Proper configuration and implementation are crucial for evaluation. In~\cite{pintor2021indicators} the authors offer guidance on preventing evasion attack failures, addressing cases where models are vulnerable yet unsuited for attack-optimization. However, there is no equivalent work regarding privacy inference attacks or backdoor detection failures. Audit results can also be compared with public benchmarks like the Robust Bench for evasion on vision models\footnote{\url{https://robustbench.github.io/}} or BackdoorBench~\cite{wu2022backdoorbench}. As before, there is a significant lack of benchmarks for non-vision tasks or other vulnerabilities.

\subsubsection{A cybersecurity audit for a system composed by several AI components}

While auditing individual AI components provides insights into their vulnerabilities, it does not guarantee an accurate approximation of the system's overall vulnerabilities. There is limited research on auditing a system comprising multiple AI components. A possible approach, inspired by the V-model from ISO26262~\cite{international2011iso} software testing, involves testing each model in isolation before integrating and testing subsequent systems. Alternatively, testing the entire system end-to-end is suggested~\cite{boloor2019simple}, with individual components only investigated upon system failure~\cite{nushi2017human}.

\subsubsection{Cybersecurity audit reports for AI components in AVs}



The output of AI components' cybersecurity audits in AVs should mirror a detailed defence evaluation~\cite{carlini2019evaluating}. This includes a rationale for chosen attacks/assessments, their parameters, and norms; benign and attack accuracies or assessment performance with associated hyper-parameters; results of basic sanity checks; and the AV components involved in the test. Evasion sanity checks, as in~\cite{carlini2019evaluating}, involve increasing attack perturbation leading to a rise in attack success rate, as well as substantial perturbation reducing model accuracy to random guessing levels. However, similar guidelines for backdoor testing or privacy assessments are currently unavailable. Privacy attack tests, though theoretically possible with overfitting leading to higher inference accuracy~\cite{liu2022ml}, are often unattainable due to the unavailability of the model owner's training data.


Audits should be conducted before AI model deployment and repeated upon significant model updates or advancements in attack techniques. To prevent model providers from tuning models to audit challenges, audit tests should not become public. Attacks can aid in developing robust models if applied during training and evaluation by ML engineers or specialized teams (e.g., AI red team). However, as stated above, to prevent overfitting to a specific attack, the audit must employ tests not used during training.

\subsection{Challenges}

Our limited understanding of AI security threats presents challenges in auditing AI components, especially within complex systems like AVs. The discrepancy between theory and practice is evident in the wealth of academic work versus the scarcity of real-world incidents, resulting in a lack of realistic threat models. This hampers our ability to test relevant attacks and understand the effects an attack might have on an individual AI component within a larger system such as an AV. This is amplified by the limited experience in auditing standalone AI systems.


Models are often optimized to pass audits, leading to 'gaming' the audit rather than ensuring robustness. Given the limited number of attacks available in libraries (Table~\ref{tab:librariesKathrin}), this is a pressing concern. The re-implementation of attacks from scientific works can introduce potential errors and inaccurate robustness assessments. This problem is amplified for non-vision based AV modules, such as LiDAR-based, with fewer existing implementations. 


While evasion attacks are most commonly implemented, followed by poisoning and IP/privacy attacks, our understanding of securing models against all these attacks remains shallow. The possibility of undetectable attacks creates a penalty-free zone for attackers, particularly when there is no significant misbehaviour. 


In terms of privacy, we need focused studies assessing privacy risks associated with AI components of AVs, including sensitive training data, understanding the use of federated learning for AVs and securing it against privacy attacks.

\section{Transparency through explainability}\label{sec5}



Explainability in AV testing and validation ensures system safety and reliability by providing an understanding of the underlying algorithms and decision-making processes. It enables performance evaluation in various scenarios and promotes transparency and accountability. As AVs become more prevalent, explaining their behavior and decisions to stakeholders and the public is crucial for building trust and confidence.

\subsection{Explainability in AVs}
In the following we provide a high-level description of the operational layers described in Section \ref{sec1} a review of previous work on explainability in each operation. In \cite{Llorca2021}, some of the most relevant barriers and questions for explainability are identified for some of these operational layers. 

\textbf{Localisation:}  
it is critical for the safe operation of AVs~\cite{wang2017map}. 
The complexity of scenarios as well as weather and traffic conditions can affect the accuracy of localisation and thereby also have an impact on the planning and decision making of a vehicle. 
Embedding a form of estimation of performance and confidence in the odometry and localisation pipelines can improve their robustness and prevent critical failures \cite{aldera_what_2019, garcia_daza_2020}. 

As safety is often considered as the most important design requirement, one of the goals of localisation is to ensure that the AV is aware whether it is within its lane~\cite{reid2019localization}. Hence, providing information about the vehicle location over time including justifications as explanations can be crucial to both exposing error rates and preventing collisions. Such explanations and/or alerts might be given through specialised user interfaces, e.g., a special dashboard or mobile application as shown in~\cite{schneider2021explain}. 

While there seems to be less research related to explainable localisation, as discussed in \cite{omeiza2022survey}, explanations concerning localisation can be useful for system developers for debugging AVs because they can facilitate positional error correction. Explanations can also provide other stakeholders with a perception of reliability and/or safety. 
\newline

\textbf{Perception:} perception and scene understanding are fundamental requirements for providing intelligible explanations. To this end, several data-driven explanation methods have been applied to the AV perception task. To train, test and validate explanation methods datasets play a critical role. 

Datasets for training machine learning approaches in the context of AVs~\cite{janai2020computer} are an important resource for explanation generation. These datasets include different types of annotations, e.g., handcrafted explanations \cite{kim2018textual,you2020traffic}, vehicle trajectories \cite{houston2020one}, human driver behaviour \cite{ramanishka2018toward,shen2020explain} or anomaly identification with bounding boxes \cite{xu2020explainable,you2020traffic}, that are helpful for posthoc driving behaviour explanation. In \cite{omeiza2022survey}, datasets have been categorised by the sensors used for the data collection (exteroception and proprioception), and the annotations that are useful for developing explainable AVs. Furthermore, different stakeholders that can potentially benefit from the explanations have been identified. 


Vision-based explanation methods represent an important class of (deep learning) methods that have been proposed. Gradient-based or backpropagation methods are generally used for explaining convolutional neural network models~\cite{das2020opportunities}. In \cite{tjoa2019survey,zablocki2021explainability} the reader can find extensive surveys on vision-based explanation methods, from Class Activation Map (CAM) \cite{zhou2016learning} and its variants \cite{selvaraju2017grad, chattopadhay2018grad, tang2019interpretable, omeiza2019smooth}, to other methods \cite{bojarski2018visualbackprop, lapuschkin2019unmasking, shrikumar2017learning}.
Furthermore, in~\cite{kim2018textual} an approach for textual explanation generation for AVs which was trained on the BDD-X dataset was proposed, and in~\cite{xu2020explainable} the authors focused on scene understanding and the generation of short explanation sequences, highlighting salient objects in input images that can potentially lead to hazards. 

As for prediction, the need to understand road users' interactions was highlighted by the authors of the SCOUT Graph Attention Network (GAT)~\cite{Carrasco2021}. They proposed the use of Integrated Gradients~\cite{ig} technique and visualization of learned attention. The illustrations provided for four carefully selected scenarios showed the importance of spatial representation of the nearest traffic participants. The use of the learned attention as a mechanism to generate explainable motion predictions was also proposed for transformer-based models \cite{Zhang2022-TRC}. 
The explainability of motion prediction was explored by use of conditional forecasting~\cite{Khandelwal2020WhatIfMP}. The authors of \textit{What if Motion Prediction} (WIMP) model~\cite{Khandelwal2020WhatIfMP} developed an iterative, graphical attention approach with interpretable geometric (actor-lane) and social (actor-actor) relationships that support the injection of counterfactual geometric targets and social contexts. The proposed method supports the study of hypothetical or unlikely scenarios, so-called counterfactuals. 
In \cite{Carrasco2023a}, the authors proposed an intention prediction layer that can be attached to any multi-modal motion prediction model, to enhance the interpretability of the outputs. The effectiveness of this approach was assessed through a survey that explores different elements in the visualisation of the multi-modal trajectories and intentions. Recently, in \cite{Carrasco2023b}, the authors propose the use of several explainability techniques integrated in a heterogeneous graph neural network model. The explainability analysis contributed to a better understanding of the relationships between dynamic and static elements in traffic scenarios, facilitating the interpretation of the results, as well as the correction of possible errors in motion prediction models.
\newline

\textbf{Planning:} in AI planning and scheduling, a sequence of actions is devised for an AV to complete a task, which helps in decision-making within the environmental dynamics~\cite{ingrand2017deliberation}. However, without explanations, the generated behaviour may confuse end-users. Explainable planning can enhance user experiences~\cite{omeiza2022survey, chakrabortiemerging}, for instance, by translating agent's plans into comprehensible formats and designing user interfaces that reinforce understanding~\cite{sado2020explainable}. Relevant works like XAI-PLAN~\cite{borgo2018towards}, WHY-PLAN~\cite{korpan2018toward}, refinement-based planning (RBP)~ \cite{bidot2010verbal}, plan explicability and predictability~\cite{zhang2017plan}, and plan explanation for model reconciliation~\cite{chakraborti2017plan,chakraborti2019plan} can provide frameworks for explainable planning in AVs.
\newline

 

\textbf{Control:} in relation to vehicle control, current ADAS interfaces now display rich digital maps~\cite{tomtom}, vehicle position, and track related attributes ahead or around the vehicle which can be considered as explanations. As discussed in \cite{omeiza2022survey}, stakeholders might be interested in demanding further explanations when the AV makes low-level decisions different from their expectations. 
\newline


\textbf{Human-Vehicle Interaction:} besides existing in-vehicle visual interfaces like mixed reality (MR) visualization \cite{sasai2015mr} and reconfigurable dashboard panels~\cite{marques2011flexible}, interfaces facilitating information exchange via voice, text, visual, gesture, or a combination of these modalities are essential. Human-vehicle interaction (HVI) plays a significant role in explaining AVs behaviours, and  its relationship to explanations is discussed further in subsequent sections.
\newline


\textbf{System Management:} Event data recorders (EDRs) in vehicles log accident-related information, aiding in post-accident fault analysis~\cite{wu2013driving}. Mandatory in US passenger vehicles since 2014, the National Transportation Safety Board (NTSB) recently highlighted the need for improved EDR requirements for ADS after the Tesla crash case in 2018~\cite{teslacrash}. As AVs proliferate, distinguishing between human driver errors and AV errors, which could stem from poor design or defects~\cite{bose2014black, kohler2014current}, is essential. These distinctions need to be clearly expressed in explanations. The current challenge of properly attributing faults to the correct traffic participant stems from difficulties in identifying and evaluating the precise cause of an accident~\cite{martinesco2019note}.

The National Highway Traffic Safety Administration (NHTSA) urges industry and standard bodies to develop uniform data recording and sharing approaches\footnote{See relevant documents here: https://www.nhtsa.gov/fmvss/event-data-recorders-edrs}. Existing inability of EDRs to provide sufficient data for crash reconstruction and liability analysis calls for solutions that can record a satisfactory number of parameters for vehicle behaviour reconstruction and explanation provision~\cite{pinter2020road}. Proposed solutions include blockchain technologies~\cite{guo2018blockchain} as well as smart~\cite{yao2020smart} and robust~\cite{HARRIS2013108} data models. With upcoming EU legislative rules on EDR~\cite{eclaw} and similar regulations in China~\cite{chinalaw}, the sufficiency of existing data storage facilities for AVs is under scrutiny. Efficient storage space management and well-defined data packages are necessary for event explanation purposes. These EDR advancements are crucial for developing accident explanation techniques, highlighting the under-explored area of explainable EDRs.

\subsection{From AV Sensor Data to Explanations}\label{sub:xai2}

\subsubsection{AV Data Recorders}

AVs generate vast amounts of data, the recording, access, analysis, and usage of which can have significant technical, ethical, social, and legal implications~\cite{TENHOLTER2022100038}. A data recording device or Black Box (BB) can enhance safety, accountability, and trust in AVs by allowing accident investigation, logging system parameters, and reconstructing incident timelines respectively. The UNECE GRVA has proposed regulatory provisions for AV-specific data recorders, the Event Data Recorder (EDR) and the Data Storage System (DSSAV). The EDR collects information on the vehicle’s behaviour and crash severity when a specific triggering event occurs (e.g. airbag activation). The DSSAV by contrast is “always on” and collects information on interactions between the driver and the system. However, these may not provide sufficient or appropriate data for practical and legal investigations, particularly in collisions or near-miss events with VRUs incidents. The ITU Focus Group on AI for Autonomous and Assisted Driving (FG-AI4AD)\footnote{https://www.itu.int/en/ITU-T/focusgroups/ai4ad/Pages/default.aspx} is assessing public expectations for self-driving software post-collision behaviour, though potential ethical risks and legal implications may provoke resistance towards AV data recording. A Responsible Innovation (RI) approach, involving societal benefit, anticipatory governance, and stakeholder inclusion considerations, can mitigate negative outcomes from inadequate data logging. The impact and challenges of RI in practice are discussed in~\cite{TENHOLTER2022100038}.

\subsubsection{Generating Explanations: Automated Commentary Driving}\label{sub:xai3}

\begin{figure}[t]
\centering
\includegraphics[width=0.8\columnwidth]{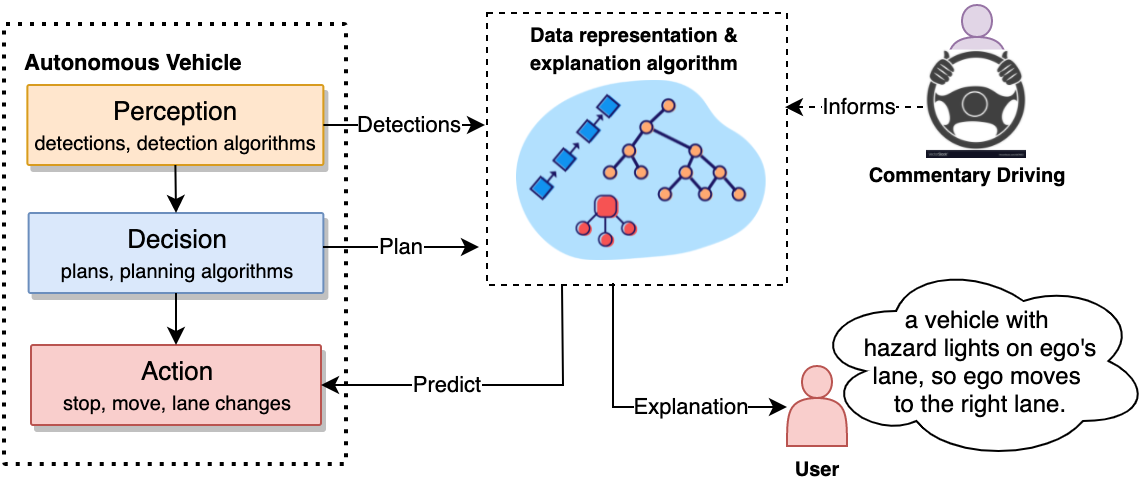}
\caption{From commentary driving, requirements for explanations were gathered to inform the design of factual and counterfactual explanation algorithms. The algorithms receive input data from the different autonomous driving operations, provide a structured representation, and generate intelligible explanations to an end-user.}
\label{fig:acd}
\end{figure}

Automated explanation generation in AVs is inspired by \emph{commentary driving}, where drivers verbalize their observations, assessments, and intentions, enhancing their understanding and awareness of their surroundings. Automated driving commentary can offer understandable explanations of driving actions, assisting drivers in challenging and safety-critical scenarios. A field study conducted in~\cite{omeiza2022acd} involved a research vehicle deployed in an urban environment to collect sensor data and driving commentary from an instructor. The analysis revealed an explanation style involving observations, plans, general remarks, and counterfactual comments. Using a transparent tree-based approach (see Figure~\ref{fig:acd}), they generated factual and counterfactual natural language explanations, with longitudinal actions (e.g., stop and move) deemed more intelligible and plausible than lateral actions by human judges. The study proposes ways to enhance robustness and effectiveness of explainability for ADAS and ADS.

\subsection{Challenges}
Despite recent and ongoing advances in the general field of explainable AI (XAI), there are a considerable number of limitations \cite{Panigutti2023} including reliability and robustness , lack of a standard evaluation framework and potential increase of automation bias. XAI is a valuable tool, but further research is needed. In the following, we summarize the main challenges for AVs. 


\subsubsection{Explaining the right thing, in the right way, at the right time for the right reason}

Determining \emph{what}, \emph{how}, \emph{when}, and \emph{why} to explain are crucial considerations for explainer system designers. These challenges, lacking trivial or unique solutions, complicate the evaluation and comparison of system-generated explanations. Hence, it is unclear how regulators would determine the minimal requirements for an explainer system. The explanation's level of detail or fidelity depends on the application domain and stakeholder, with different groups requiring varied explanation types. For instance, accident investigators require more technical details than end-users. The manner and timing of explanations are also essential, with human factors playing a significant role. Some explanations aim to enhance understanding and trust, while others might be safety-critical. Ideally, explainer systems should adjust to the varying needs and requirements of the domain and stakeholders. This challenge calls for research into new algorithmic methods for generating effective explanations from AV sensor data and plans, clear standards or regulations on what and why AVs should explain (e.g., see \cite{winfield2022open}), and further human factors research to optimize explanation provision.

\subsubsection{Explaining the unknown unknowns}



AVs operating in open-ended environments will face unfamiliar situations or edge cases, potentially making it challenging to generate appropriate explanations. For instance, an AV may encounter an unclassifiable new type of road user, making explanations based on incorrect interpretations potentially misleading for stakeholders. To tackle this, AVs require meta-cognition or reasoning capabilities to assess if any functionality is compromised or uncertain. This necessitates new meta-reasoning methods and techniques for explaining this new type of inference to stakeholders.

\subsubsection{Explaining opaque AI models}


Deep learning methods have significantly advanced the field of computer vision and are now fundamental in many applications, including AVs. Despite their high-performance, they lack interpretability and explainability due to their size, structure, and parameter count. Progress in XAI has driven research to make neural networks understandable, with many approaches focusing on vision-based explanations using attention. Although useful for building an intuitive understanding of otherwise opaque AI models, these explanations have limited explanatory power. Therefore, new approaches that can provide robust causal explanations are required.

\section{Robustness and Fairness in AVs}\label{sec6}
This section examines the safety and ethical behaviour of AVs. We will discuss potential evaluation protocols and metrics for assessing predictive system robustness, and consider a strategy for implementing ethical objectives to guide AV behaviour towards fairness-based goals.

\subsection{Evaluating Predictive System Robustness in AVs}

As ADAS and ADS increase in complexity, conventional testing struggles to address all safety areas due to the vast array of safety-related systems and potential scenarios. The prediction pillar of the perception operational layer is crucial in forecasting agent-agent and agent-scene interactions, with models typically representing multi-modal paths to account for environmental uncertainty, adding complexity to evaluation criteria~\cite{Carrasco2023a}. Following is an overview of the evaluation metrics used to validate observed ground truth predictions, along with a discussion on the application of robustness analysis strategies.
\newline


\textbf{Unimodal metrics:} they evaluate single predictions using distance-based metrics, including Average Displacement Error (ADE), Final Displacement Error (FDE), and Heading Error. However, due to the multi-modal nature of human behaviour, metrics should evaluate multiple plausible outputs.
\newline

\textbf{Multimodal metrics:} measure the likelihood of all possible futures given past observations. Common metrics include Top-K, which samples K predictions and returns the best prediction performance based on a distance measure. Despite its popularity, Top-K is potentially misleading and fails to measure the social plausibility of all modes (e.g., the collision rate of each prediction). The ideal metric should also be able to quantify that the model does not generate bad trajectories Other metrics such as Average NLL~\cite{Trajectron2019} and Average Mahalanobis distance (AMD)~\cite{Mohamed2022} try to measure the likelihood of ground truth and variance/certainty of predicted outputs, respectively. The Percentage of Trajectory Usage~\cite{Li2022} is a promising metric that captures accuracy, improving on Top-K. However, developing metrics that holistically capture multimodality remains an open issue \cite{Carrasco2023a}.
\newline

\textbf{Socially-aware metrics:} measure social awareness, for example, by counting collisions in model outputs or assessing overlap rate~\cite{Carrasco2023a}.
\newline

\textbf{Scene-aware metrics:} evaluate scene awareness, measuring variables such as the \textit{number of off-roads}, Drivable Area Occupancy (DAO), Drivable Area Compliance (DAC), Scene Consistency Rate (SCR), or Oncoming Traffic Direction (OTD)~ \cite{Carrasco2023a}.
\newline


\textbf{Imbalanced datasets:} are common in human mobility, with critical non-linear behaviours being rare (most of the data is repetitive linear behaviours). Even with millions of examples, not all behaviours may be observed in the training set. Thus, models should be evaluated on a diverse set of interactions to assess robustness. One approach is to categorize data by interaction type (e.g., static, linear, leader-follower, collision avoidance, road following, etc.) and evaluate each category independently~\cite{trajnet++}. Another strategy is to develop methods that automatically identify \textit{realistic adversarial examples} that cause model failure.
\newline

\begin{figure*}[t]
  \centering
  \begin{subfigure}[b]{0.24\linewidth}
    \centering\includegraphics[width=\linewidth]{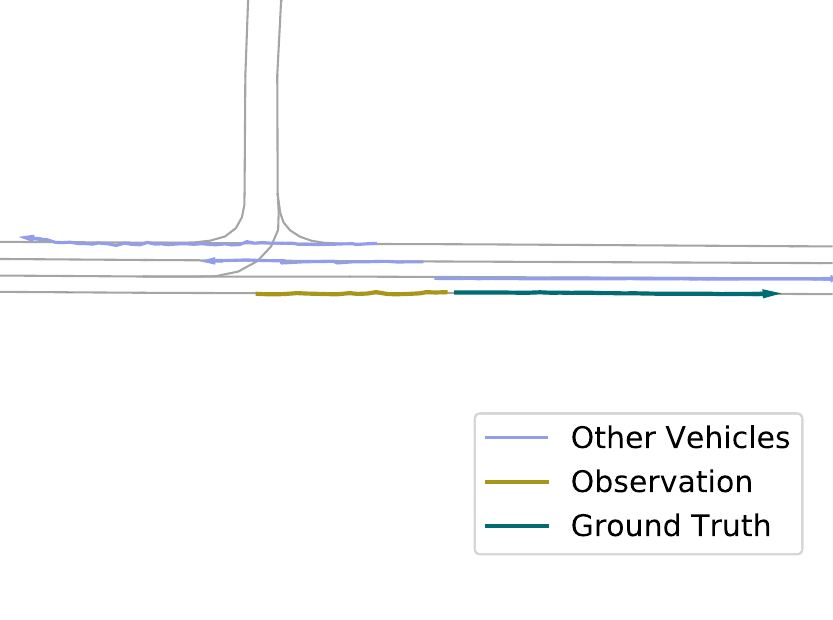}
    \caption{Before transformation}
  \end{subfigure}%
  \hfill 
    \begin{subfigure}[b]{0.24\linewidth}
    \centering\includegraphics[width=\linewidth]{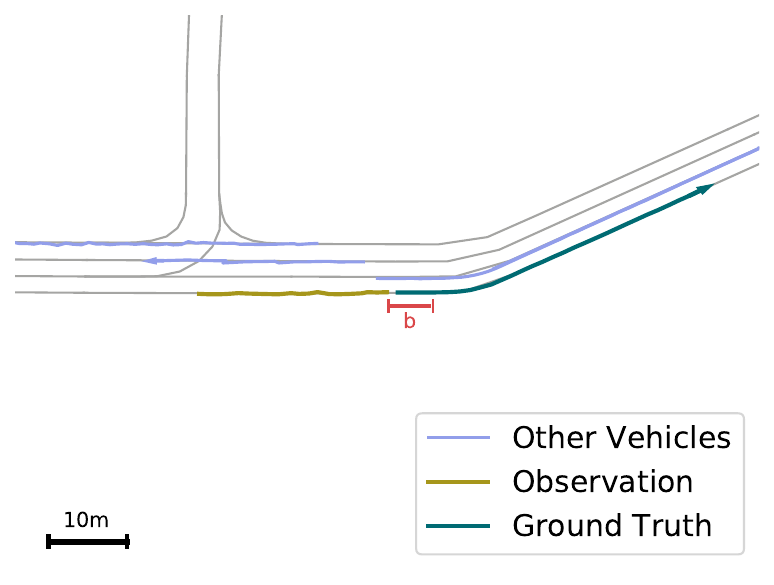}
    \caption{Single-turn}
  \end{subfigure}%
  \hfill
    \begin{subfigure}[b]{0.24\linewidth}
    \centering\includegraphics[width=\linewidth]{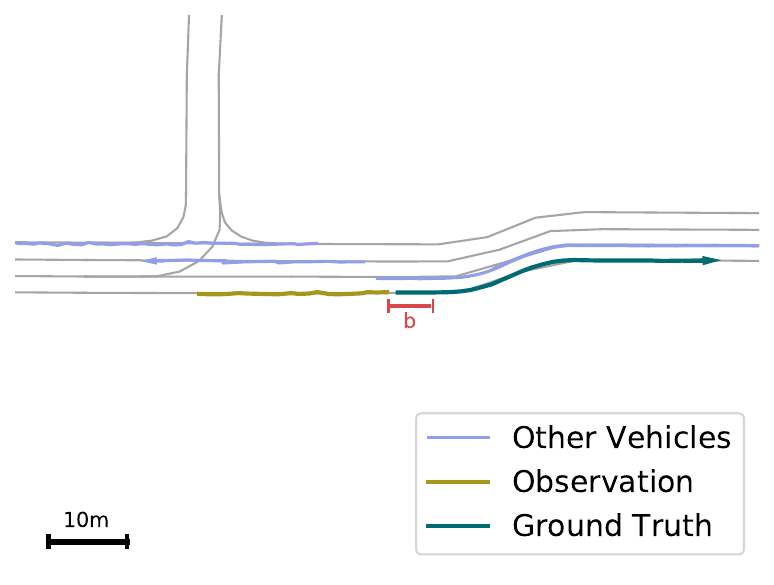}
    \caption{Double-turn}
  \end{subfigure}
  \hfill
    \begin{subfigure}[b]{0.24\linewidth}
    \centering\includegraphics[width=\linewidth]{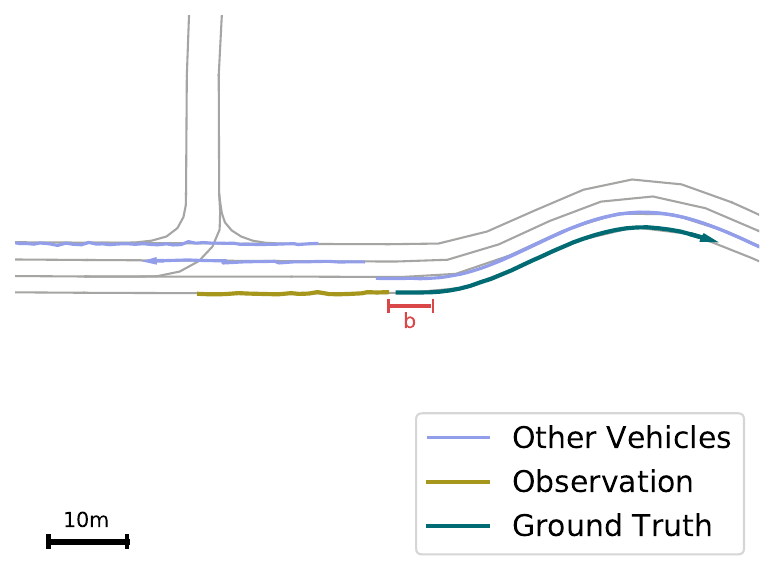}
    \caption{Ripple-road}
  \end{subfigure}
  \caption{Visualization of different transformation functions. The scene before transformation is followed by three different transformations.}
  \label{fig:attack_methods}
\end{figure*}

\begin{figure*}[t]
 \centering
    \centering    \includegraphics[width=0.8\linewidth]{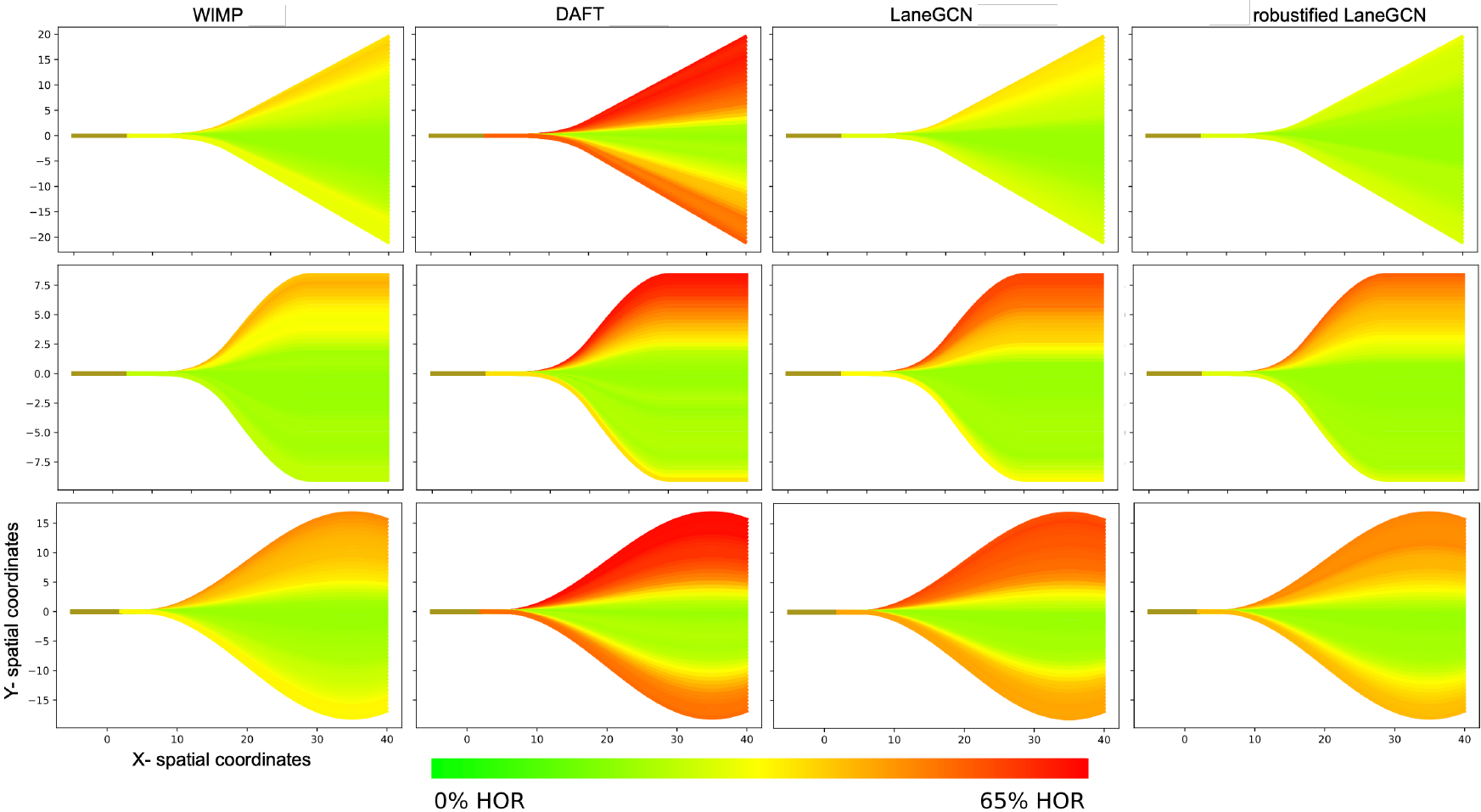}
 \caption{The qualitative results of various methods (WIMP~\cite{khandelwal2020whatif},DAFT~\cite{park2020diverse}, LaneGCN~\cite{liang2020lanegcn}) for different transformation functions. 
 The red color indicates high Hard Off-road Rate (HOR) in those scenes.}
 \label{fig:cool_exp}
\end{figure*}

\begin{figure*}[t]
  \centering
  \begin{subfigure}[b]{0.327\linewidth}
    \centering\includegraphics[width=\linewidth]{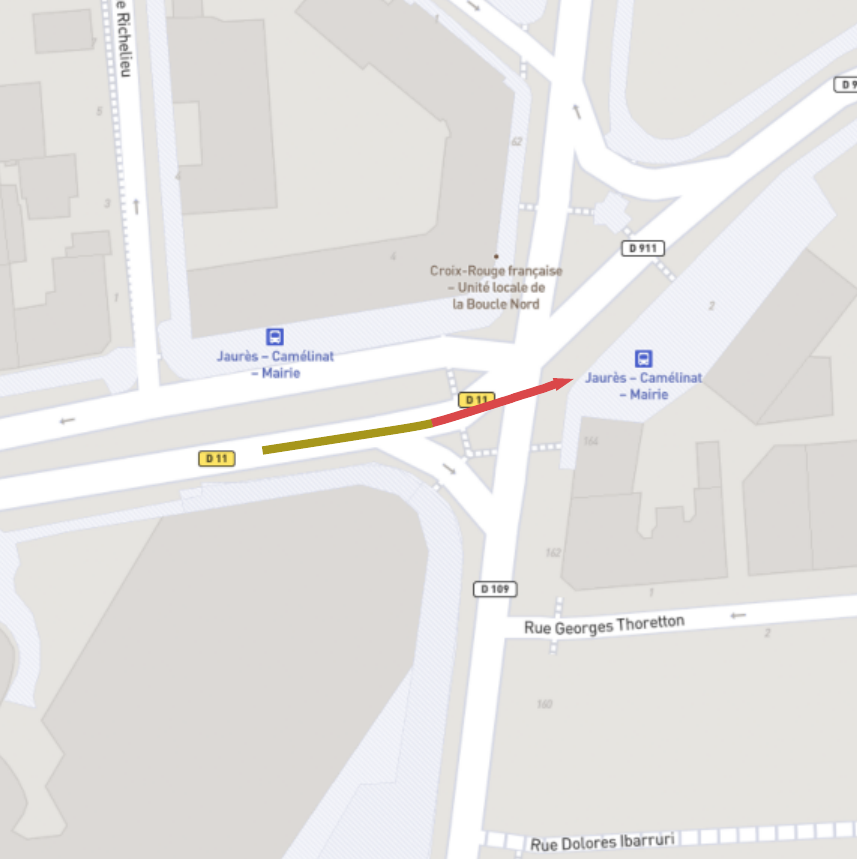}
    \caption{Paris \href{https://www.google.com/maps/@48.9267618,2.2944925,19z}{location}}
  \end{subfigure}%
  \hfill
  \begin{subfigure}[b]{0.327\linewidth}
    \centering\includegraphics[width=\linewidth]{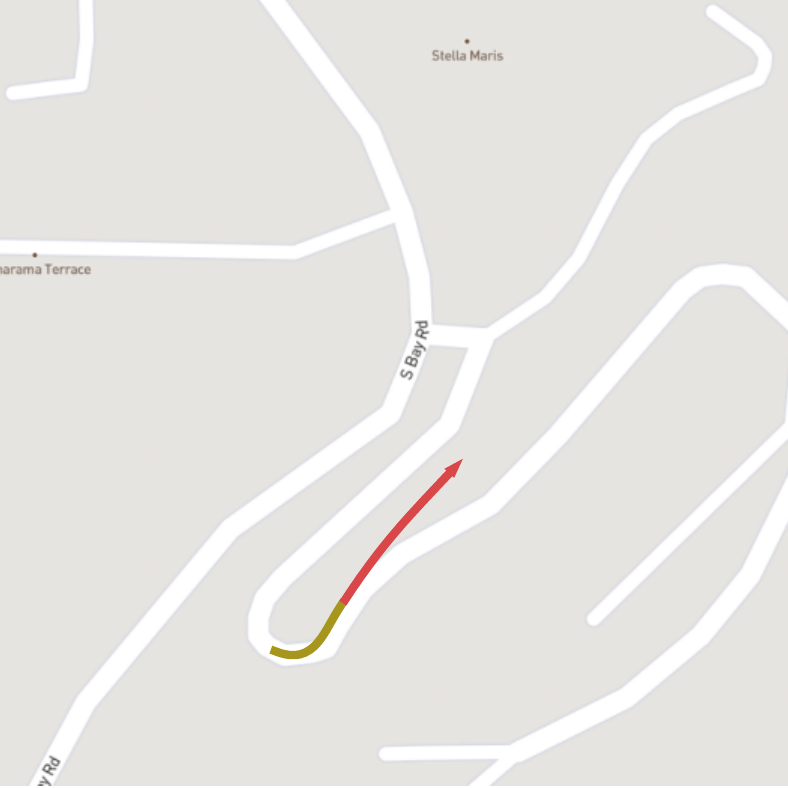}
    \caption{Hong Kong \href{https://www.google.com/maps/@22.2332883,114.1992906,20.25z}{location}}
  \end{subfigure}
  \hfill
    \begin{subfigure}[b]{0.327\linewidth}
    \centering\includegraphics[width=\linewidth]{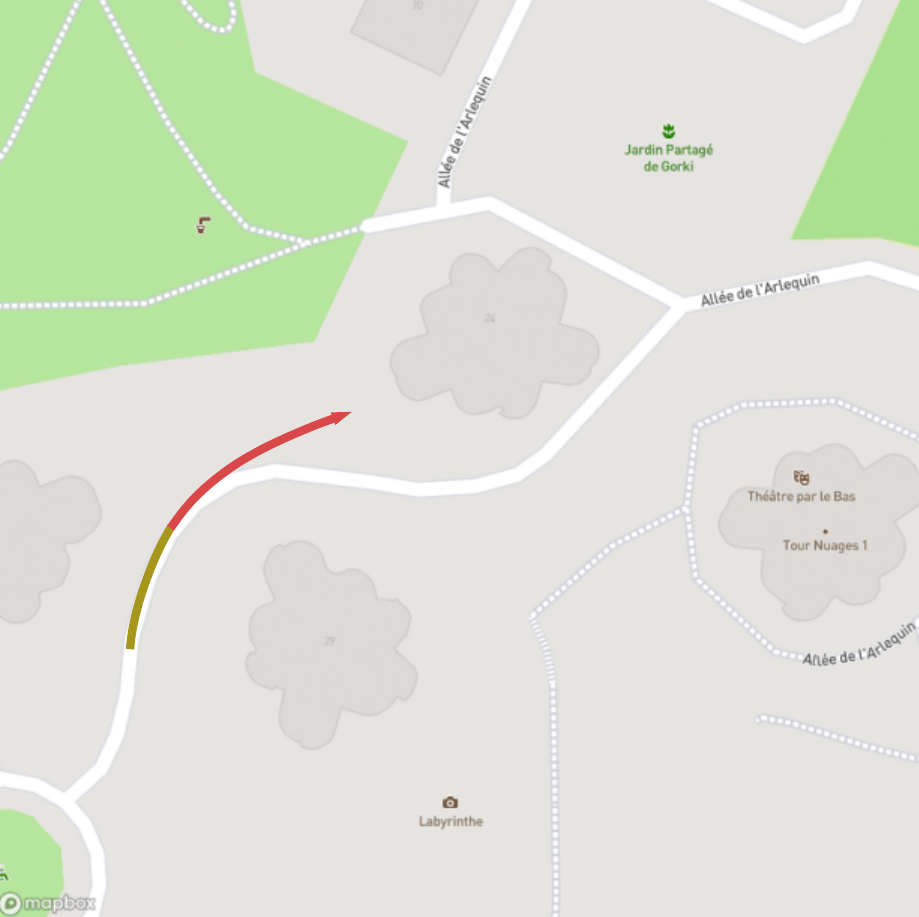}
    \caption{New Mexico \href{https://www.google.com/maps/@48.8905404,2.2268305,19.25z}{location}}
  \end{subfigure}
  \caption{
  It is possible to automatically find more models fail without collecting data from these real-world scenes. The automatically generated scenes are used to retrieve real scenes. We can observe that a model fails in Paris (a), Hong Kong (b) and New Mexico (c) in a few milli-seconds.}
 \label{fig:real-world}
\end{figure*}

\textbf{Robustness assessment through adversarial examples:} involves identifying realistic examples that cause prediction models to fail. Such examples could be specific interactions not seen in the training data. The task is to automatically find a realistic modification of the scene and agents' positions that cause prediction models to fail. Off-road predictions, for example, indicate a failure in the model's scene reasoning~\cite{park2020diverse}. To find real examples where the models go off-road, the space of possible scenes should be explored either by learning generative models that mimic the dataset distribution or by using simple functions to transform the scene into new, realistic but challenging ones. Figure \ref{fig:attack_methods} illustrates how a scene can be automatically altered into a realistic example for evaluation. We can then quantitatively report the frequency of model failures in specific scenes (as shown in Figure \ref{fig:cool_exp}). Lastly, we can verify the plausibility or existence of the generated failing scenes. We can find a similar real scene using image-retrieval techniques, given a generated scene. Figure \ref{fig:real-world} depicts automatic retrieval of real-world scenes where models fail without the need for data collection from these scenes. 

Similarly, changing agents' positions can help identify realistic interactions where models fail, such as predicting a collision. The sensitivity of various models in different timesteps with regard to collision avoidance can be studied~\cite{Saadatnejad2022}, which might be explained by biases in the data or in the model structure.

\begin{figure}[t]
    \begin{center}
        \includegraphics[width=0.7\columnwidth]{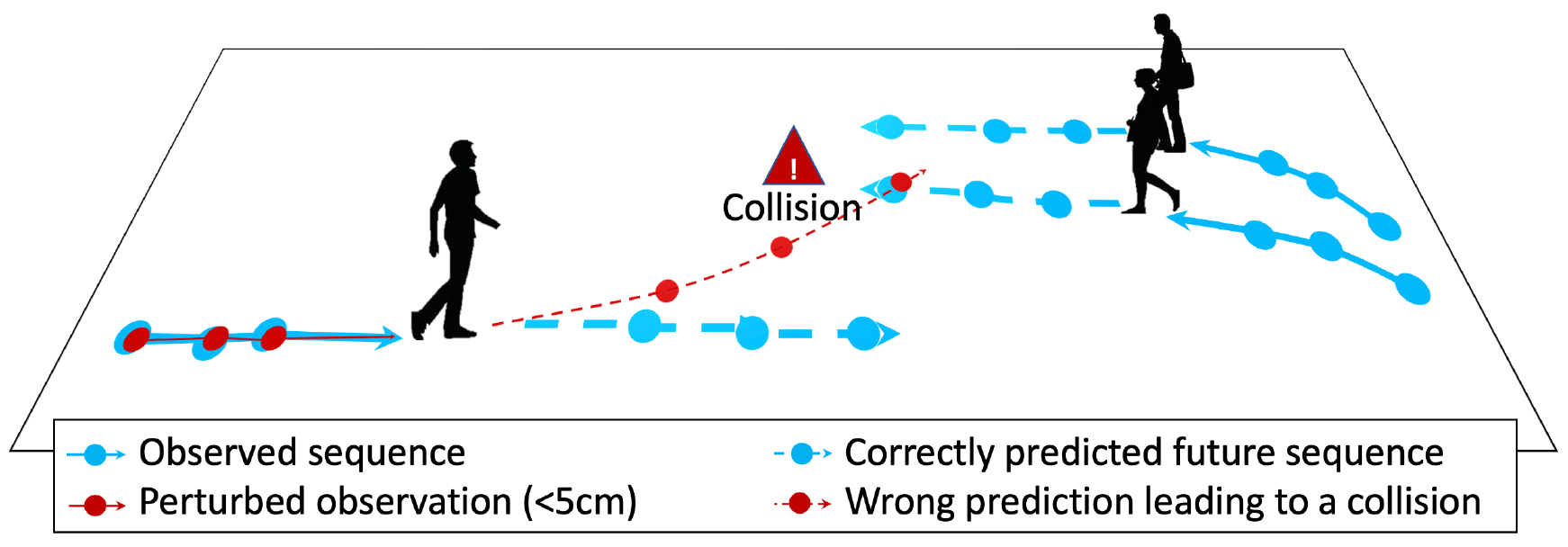}
    \end{center}
    \caption{Given the observed trajectories of the agents in the scene, a predictor forecasts the future positions reasonably (blue lines). However, with less than $5$~cm perturbation in the observation trajectory (in red), an unacceptable collision is forecasted. }
\label{fig:pull}
\end{figure}

Finally, while it is possible to identify agents' social interactions that could lead to a prediction failure, the robustness of small, realistic perturbations to the observed data can also be studied. For example, a model might accurately predict future positions without collisions based on observed human trajectories. However, introducing a small perturbation of less than 5 cm to the observation trajectory could unexpectedly result in a collision between agent predictions, indicating incomplete social understanding by the predictors (as seen in Figure \ref{fig:pull}). Unlike common adversarial attacks designed for classifiers\cite{NEURIPS2020_1ea97de8}, it is possible to design attacks for the trajectory prediction problem, which are a multimodal regression task. Realistic adversarial examples can be used to study the collision avoidance behaviour of trajectory prediction models. Specifically, we can investigate the worst-case social behaviour of a prediction model under small input perturbations, including various levels of ablation for agents and scene elements~\cite{Carrasco2023b}.

\subsection{Ethical goals}
The implementation of vehicle automation is anticipated to improve road safety by eliminating common human driver errors~\cite{Morando2017Investigating, Lu2016Human}. However, new risks may arise from factors like mode confusion, loss of situational awareness, and control transition problems~\cite{Carsten2019How, Cummings2014Shared, Martens2013road}. Additionally, the infinite variety of situations that AVs may encounter could lead to new types of collision risks.

A report for the European Commission~\cite{Bonnefon2020Ethics} included twenty recommendations addressing the ethics of AVs, covering topics such as road safety, privacy, fairness, explainability, and responsibility. The recommendations emphasized that AVs must decrease, or at least do not increase physical harm to users or other road users compared to conventional driving. Furthermore, even if AVs reduce the global level of risk, their deployment would be considered unethical if any individual group of road users experienced an increase in risk. The introduction of AVs also requires careful consideration of circumstances in which they might be permitted not to comply with traffic rules.

Current road rules use terminology that allows discretion by human drivers, such as driving with ‘reasonable consideration’ and ‘due care’. Human drivers exercise this discretion based on experience, training and a general understanding of the road environment. AVs could be designed to follow digital versions of statutory road traffic rules. However, they may face challenges conforming to traffic law due to the complexity of their operating domains~\cite{Prakken2017problem}. The complexity of domains in which AVs are expected to operate makes it impossible to anticipate or formulate explicit rules for every potential eventuality - environmental conditions, traffic and encounters with other road users will vary dramatically between domains and over time in any one domain. Any approach that assumes an AV could "deduce" normatively correct behaviour through exposure to a large number of training cases would need to overcome three extremely challenging practical difficulties. First, collecting a sufficient quantity and quality of scenarios to allow the right behaviours to be derived. Second, to derive the ethical principles underpinning which decisions or behaviours should be adopted. And third, to explain or justify its decisions or actions. 

An alternative solution explored in~\cite{Reed2021Ethics} is the potential use of ethical goal functions, which capture societal expectations on how an AI should optimise outcomes. For an AV, this would involve setting out priorities for safe and efficient driving. Ethical goal functions could guide decisions where non-compliance with road rules is preferable and align AV behaviours with societal expectations.

Once an ethical goal function has been agreed and enacted by legislators, AVs could use it to determine the course of action with the highest utility. Factors considered within the ethical goal function to determine optimal speed, for example, might include potential collision risk, road suitability, infrastructure, vehicle and weather conditions, and the behaviour of other road users. A similar calculation that incorporates this function could assess whether mounting the pavement is permissible, in what circumstances and how that operation should be performed.

Ethical goal functions offer a means to enhance societal engagement and assist developers in AV deployment by providing algorithmic guidance on acceptable behaviours. Demonstrating how AVs have optimized against ethical goal functions could help show that their vehicles act in accordance with societal values, facilitating any possible liability claim process~\cite{liabAI}. Also, it gives transport regulators an effective tool to monitor and moderate AV behaviours.

\subsection{Challenges}
\subsubsection{Robustness of Prediction}

Thoroughly assessing forecasting models necessitates extensive data collection, which is often impractical and costly. It is unrealistic to assume that all possible social interactions can be observed in data collection campaigns. Recent research suggests new testing and validation protocols using both real and synthetic data to investigate trajectory prediction model robustness~\cite{bahari2021vehicle,saadatnejad2021sattack}. Another approach introduces real subject behaviour in autonomous driving simulators using virtual reality and motion capture systems~\cite{MartinSerrano2022, MartinSerrano2023a, MartinSerrano2023b}. More broadly, current procedures for testing prediction systems could benefit from improved fidelity (e.g., more realistic dummies), increased variability of testing conditions (e.g., dummies with different appearances, partially changing trajectories), and the addition of real behaviours and interactions in a safe, controlled manner (e.g., using simulated and interactive environments)~\cite{Izquierdo2022}.

\subsubsection{Encoding ethical behaviour in AVs}
While AV deployment has taken longer than was anticipated, it has given time for researchers and society to consider exactly how we might expect AVs to behave and how we want them to integrate into society. Application of ethical goal functions could ensure that society and human values steer the operation of AVs and help to ensure that when the technology arrives, it delivers on societal expectations. Work is required to develop the process for capturing societal values and to set out the framework by which these values are successfully translated and encoded into AV behaviours but this is work that may be on the critical path to ensuring that AVs drive safely and ethically.

\section{Conclusion}\label{sec7}
AI is pivotal in the evolution of AVs, leading to high levels of automation. Despite the benefits, AI-specific features such as unpredictability, opacity, self and continuous learning, and lack of causality can introduce new uncertainties and safety issues. Recognizing this, international regulatory bodies are addressing the impact of AI on vehicle regulations. To harness the benefits of AI while mitigating potential risks, several frameworks have been proposed to guide the development of trustworthy AI systems, incorporating principles such as cybersecurity, robustness, fairness, transparency, privacy, accountability, and societal and environmental well-being. These frameworks present complex, varied challenges requiring multidisciplinary expertise. In this work, based on an incremental methodology that coordinates the work of a multi-disciplinary group of experts, we have presented a detailed analysis of the state of the art in cybersecurity, transparency, robustness, and fairness in the context of AV testing procedures, identifying future challenges.

We have discussed the role of AI in AVs, the challenges in their development, testing, regulation, and accident investigations, the need for cybersecurity audits, and the importance of transparency and explainability. AI plays a major role in AVs' functioning and their interaction with the environment. The EU’s AI Act could influence AVs, but clarity is needed on its application, especially regarding AI components as safety components. The development of AVs is guided by the V-model and influenced by vehicle regulations, but current testing methodologies face limitations due to increasing complexity. Post-market surveillance and in-service monitoring are vital for ongoing safety assessment. Accident investigations involving AVs are complex and may benefit from lessons learned in aviation. 

On the security front, structured cybersecurity audits for AI components in AVs are necessary to prevent threats. We propose a four-level framework for AI security research. Regular audits, especially before production and after significant updates, are advised. However, challenges in auditing include lack of realistic threat models, absence of real-world incidents, and potential for models to be optimized for passing audits. Transparency and explainability are key to understanding AVs' decision-making processes. This requires domain-specific explanations and understanding of the datasets used to train machine learning models. Challenges include determining what to explain and how, dealing with unknown scenarios, and explaining opaque AI models. Lastly, the paper discusses evaluation protocols for assessing predictive system robustness and ethical behaviour in AVs. Different metrics are used to evaluate predictive system robustness, but a holistic approach is needed. Additionally, the use of ethical goal functions to capture societal expectations for AV operation is considered. Challenges include the extensive data collection required for model assessment and the complexity of encoding ethical behaviour in AVs.

\backmatter

\bmhead{Disclaimer}
The views expressed in this article are purely those of the authors and may not, under any circumstances, be regarded as an official position of the European Commission.

\bibliography{sn-bibliography}

\end{document}